\documentclass[11pt]{article}

\usepackage[final]{acl}

\usepackage{times}
\usepackage{latexsym}
\usepackage[T1]{fontenc}
\usepackage[utf8]{inputenc}
\usepackage{microtype}
\usepackage{inconsolata}
\usepackage{graphicx}
\usepackage{adjustbox}
\usepackage{array}
\usepackage{multirow}
\usepackage{bm}
\usepackage{amsmath,amsfonts}
\usepackage{tikz}
\usepackage{booktabs} 
\usepackage{times} 
\usepackage{helvet}  
\usepackage{courier}  
\usepackage{algorithm}
\usepackage{algorithmic}
\usepackage{natbib}  
\usepackage{caption} 
\usepackage{makecell}  
\usepackage{graphicx} 
\usepackage[table]{xcolor}
\usepackage{enumitem}

\newcommand*\bigcircled[2][1]{\tikz[baseline=(char.base)]{ \node[draw, circle, inner sep=0.7pt, scale=#1] (char) {#2};}}

\title{ChatHLS: Towards Systematic Design Automation and Optimization for High-Level Synthesis}

\author{
 \textbf{Runkai Li\textsuperscript{\rm \dag 1,2}},
 \textbf{Jia Xiong\textsuperscript{\rm \dag 1,2}},
 \textbf{Xiuyuan He\textsuperscript{\rm 2}},
 \textbf{Jieru Zhao\textsuperscript{\rm 3}},
 \textbf{Jiaqi Lv\textsuperscript{\rm 1}},
 \\
 \textbf{Haowen Fang\textsuperscript{\rm 2}},
 \textbf{Lei Qi\textsuperscript{\rm 1}},
 \textbf{Xi Wang\textsuperscript{\rm $\ast$1,2}}
\\
 \textsuperscript{\rm 1}Southeast University, Nanjing, China
 \\
 \textsuperscript{\rm 2}National Center of Technology Innovation for EDA, Nanjing, China
 \\
 \textsuperscript{\rm 3}Shanghai Jiao Tong University, Shanghai, China
}

\begin{document}
\maketitle

\begingroup
\renewcommand\thefootnote{} 
\footnotetext{\textsuperscript{\dag} \small{Equal Contribution.}}
\footnotetext{\textsuperscript{$\ast$} \small{Corresponding Author: \href{mailto:xi.wang@seu.edu.cn}{xi.wang@seu.edu.cn}}}
\addtocounter{footnote}{-2} 
\endgroup

\begin{abstract}

High-Level Synthesis (HLS) improves IC development productivity by enabling hardware design from C-like languages. However, strict coding constraints and design-specific optimizations limit its widespread adoption. While recent efforts employ large language models (LLMs) to assist HLS design, they often struggle with synthesizability rules and directive semantics.
To this end, we introduce ChatHLS, a multi-agent HLS design framework that leverages specialized LLMs for automated debugging and directive tuning. ChatHLS incorporates an adaptive error case expansion mechanism, combined with a reasoning-to-instruction analysis method to accurately diagnose HLS errors. To optimize hardware performance, it enables QoR-aware reasoning to learn the impact of HLS directives on the quality of results (QoR). Experimental results demonstrate that ChatHLS outperforms Gemini-3-pro with a 32.6\% relative improvement in debugging, while achieving significant speedups across various HLS kernels and neural network accelerators. These results underscore the potential of ChatHLS for agile hardware development.
\end{abstract}

\section{Introduction}
\label{sec:Introduction}

\begin{figure}[t]
    \centering
\includegraphics[width=7.8cm]{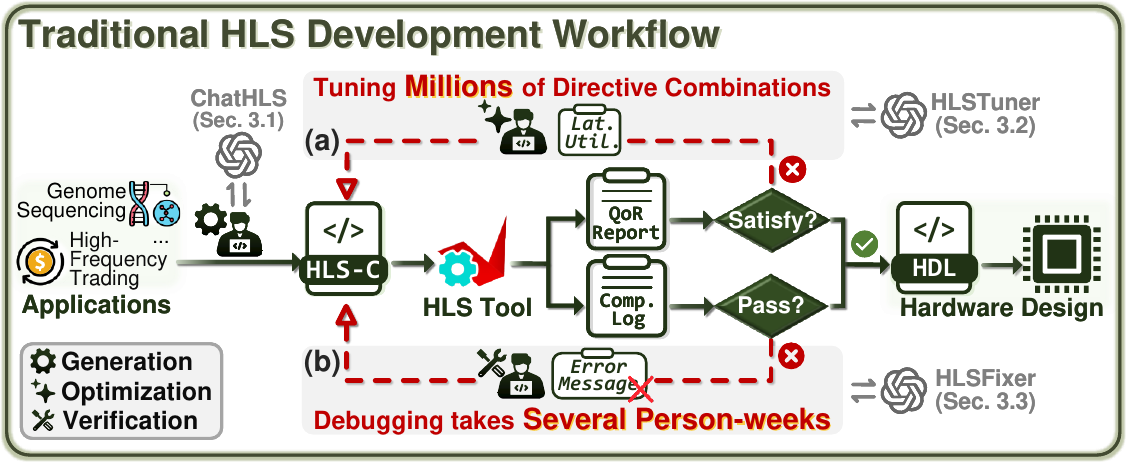} 
\vspace{-20pt}
    \caption{Traditional HLS generation, optimization, and verification workflow. Bottlenecks are: (a) balancing performance and resource utilization requires time-consuming directive tuning and (b) fixing simulation and synthesis errors relies on HLS domain expertise.}
	\label{motivation}
    \vspace{-15pt}
\end{figure}

High-Level Synthesis (HLS) accelerates hardware design by abstracting hardware description languages (HDLs) to C/C++ \cite{FPGAHLSToday}. By allowing designers to focus on algorithmic logic rather than cycle-by-cycle circuit behavior, HLS enables rapid development with reduced coding complexity and shorter simulation cycles \cite{allo}. This design agility enables rapid iteration in demanding applications such as deep learning and high-frequency trading \cite{Democratizing}. Although HLS improves hardware development efficiency, its practical adoption is severely stymied by a prohibitive design space exploration (DSE) challenge. Achieving optimal performance requires navigating an exponentially large search space of directive combinations (e.g., loop unrolling, pipelining, and tiling factors). These directives often exhibit complex interdependencies, where a single parameter change can lead to drastic fluctuations in resource utilization and throughput. Crucially, the iterative directive tuning is hindered by high synthesis latency, as a single trial often consumes minutes to hours. This creates a bottleneck of exhaustive trial-and-error, making it nearly impossible for manual designers to reach the optimal HLS design within a reasonable development cycle (Figure \ref{motivation}).

\begin{figure}[t]
    \centering
	\includegraphics[width=7.5cm]{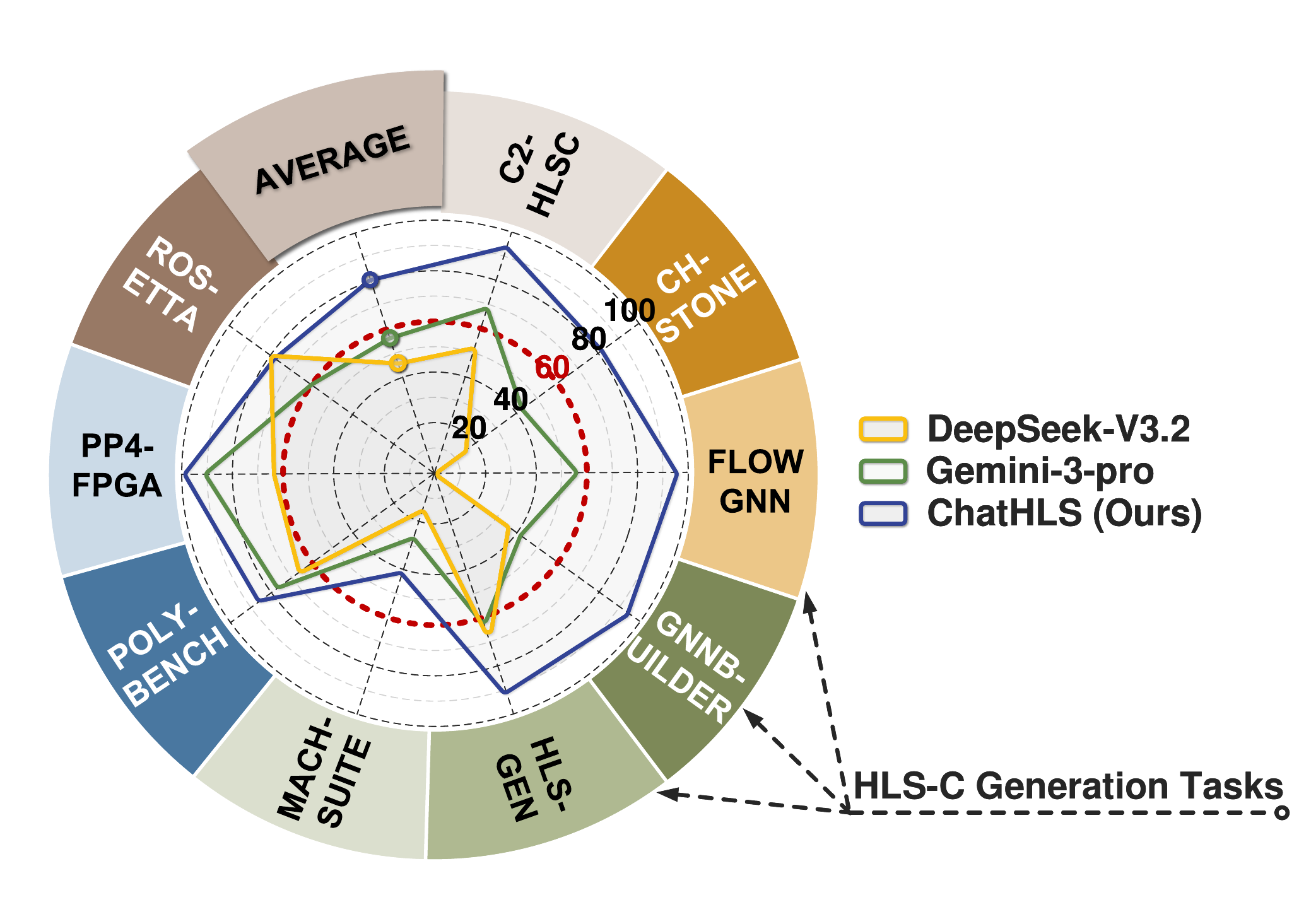} 
    \caption{Average simulation pass rates of existing LLMs generating HLS-C from natural language algorithm descriptions \cite{hlseval}, calculated across 108 tasks and averaged over 20 repetitions (see Appendix \ref{appendix:HLSFixer_Setting} for detailed experimental settings).}
	\label{Background}
    \vspace{-5pt}
\end{figure}

The success of large language models (LLMs) in mainstream programming languages has inspired LLM-aided design (LAD) in IC design and verification \cite{ChatCPU,chipmind,chatsva, fixme}. This trend provides a novel solution to address the challenges in HLS development. Recent efforts have introduced LLMs to automate the refactoring of the C/C++ code into HLS-C and the insertion of directives for hardware optimization \cite{HLSPilot, C2HLSC, ralad}. While these approaches explore the feasibility, the scarcity of HLS-specific datasets limits the correctness of generated HLS-C and the effectiveness of performance optimization \cite{hlsdebugger}. Specifically, we identify three critical bottlenecks:

\textbf{Challenge 1: HLS Data Scarcity.} 
Constructing high-quality HLS datasets is extremely labor-intensive, relying on manual construction by domain experts \cite{hlsrewriter}. More critically, existing datasets rarely expose synthesizability constraints, the rationale behind directive selections, and their correlation with quality of results (QoR) (e.g., latency and resource usage). This scarcity hinders LLMs from learning hardware constraints and the intricate semantics of HLS directives. Consequently, the generated code frequently suffers from compatibility errors, as evidenced by pass rates generally \textbf{below 60\%} shown in Figure \ref{Background}.

\textbf{Challenge 2: Inefficient Performance Optimization.} 
HLS optimization suffers from a combinatorial explosion of directive choices, whose QoR effects are highly non-linear and design-dependent. Consequently, directive tuning becomes inherently time-consuming.
Existing LLMs lack the architectural intuition required for design-specific tuning, struggling to determine the optimal combination, configuration, and insertion of directives, resulting in suboptimal hardware performance.

\textbf{Challenge 3: Limited HLS-C debugging capabilities.} Since HLS prohibits non-synthesizable constructs (e.g., dynamic arrays), general-purpose LLMs, which are pretrained on large-scale standard C/C++ datasets, struggle to identify and correct these HLS compatibility errors. Meanwhile, invalid optimizations or synthesis errors caused by incorrect directive syntax, placement, or functional conflicts further limit their debugging ability.

To address the challenges in LLM-driven HLS design, we propose \textbf{ChatHLS}, a multi-agent framework for agile HLS-C generation and optimization. ChatHLS incorporates a verification dataset construction method based on dual-agent collaboration, directing LLMs toward more comprehensive HLS error correction. By analyzing verification feedback from HLS tools, we extract expert-like debugging reasoning patterns to augment LLMs for reasonable error diagnosis. To achieve effective HLS design optimization, ChatHLS learns to perform \textit{QoR-aware reasoning}. This enables automated directive tuning, striking a balance between performance gains and resource consumption. 
Our contributions are summarized as follows:

\begin{itemize}[leftmargin=*, nosep]

\item We propose \textbf{HLSTuner}, a \textit{QoR-aware directive optimization framework} that explicitly models the directive $\rightarrow$ synthesized hardware $\rightarrow$ resulting QoR and performs constraint-aware search to navigate performance-cost trade-offs.

\item We introduce \textbf{HLSFixer}, a \textit{hierarchical feedback-augmented debugging framework} that grounds LLM reasoning in tool feedback. HLSFixer formulates debugging as error diagnosis and correction instruction, using a reasoning-to-instruction procedure to correct HLS-specific errors.

\item We propose \textbf{Verification-Oriented Data Augmentation (VODA)}, a \textit{self-evolving error case expansion mechanism} that automates the capture of error cases detected in HLS design to strengthen the debugging capabilities of LLMs.

\item Experimental results show that ChatHLS outperforms Gemini-3-pro, improving the HLS-C generation success rate by 41.8\% and error analysis accuracy by 32.6\%, while achieving a\textbf{ 3.3$\bm\times$} performance gain over the RAG-based method.

\end{itemize}

\section{Related Work}
\label{sec:Background}

\begin{figure*}[t]
	\centering 
    \includegraphics[width=16cm]{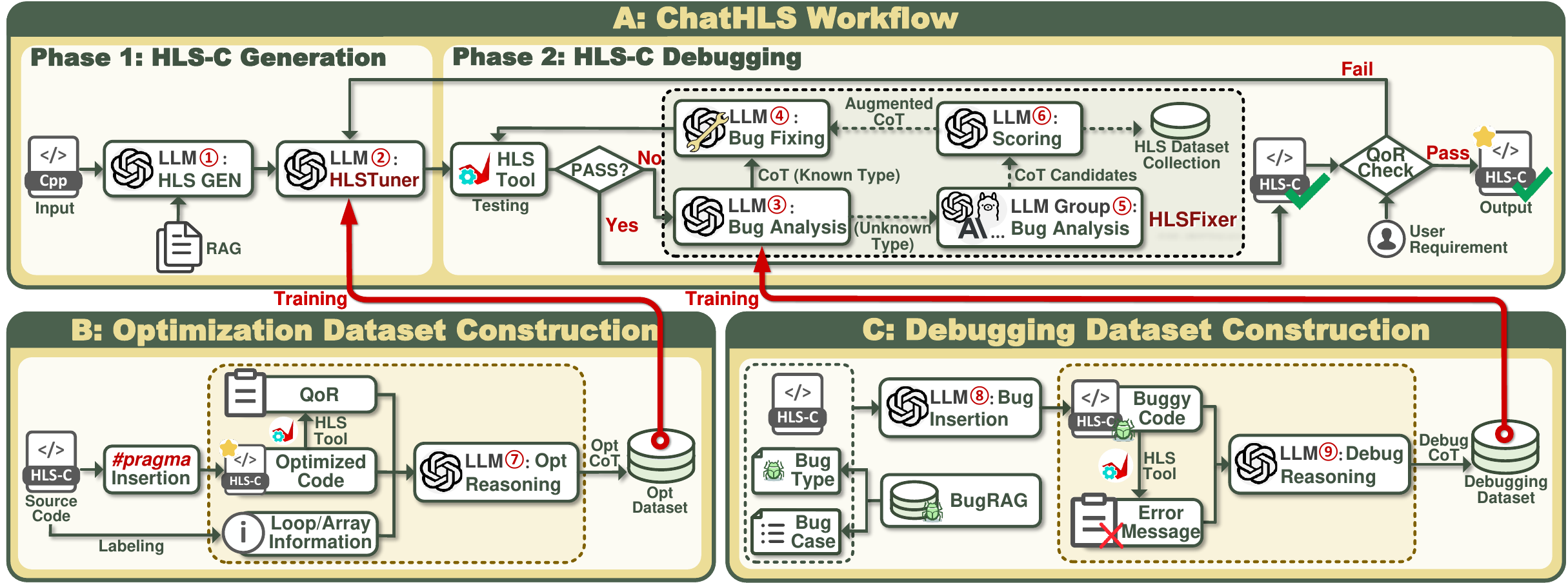} 
	\caption{ChatHLS workflow and dataset construction.}
	\label{workflow}
    \vspace{-8pt}
\end{figure*}

\subsection{Traditional Alignment to HLS Design}
Unlike C/C++ programming, HLS developers must refactor C code to align with HLS paradigms. HeteroRefactor automates the refactoring of C to HLS-C through dynamic invariant analysis \cite{heterorefactor}. HeteroGen advances this approach with fuzzing tests for automated test input generation and exception handling \cite{HeteroGen}. However, both solutions require predefined templates and manual oversight to ensure the synthesizability. Domain-specific languages (DSLs) further abstract algorithmic representations and hardware optimization in HLS \cite{allo, Dahlia, HeteroCL, scalehls}. While DSLs mitigate certain coding pitfalls, they introduce additional learning curves and exhibit limited expressivity, restricting applicability to nuanced use cases. 

To achieve satisfactory synthesized hardware performance, HLS designers also need to consider the quality of results (QoR) and strategically apply hardware-specific directives.
However, the combinatorial explosion of directives constitutes an overwhelming design space \cite{DSEReview}. Traditional methods rely on heuristics \cite{AutoDSE} or prediction models \cite{hgbo_dse, moe_dse} to identify optimal combinations and configurations of directives. Nevertheless, heuristic-based approaches require numerous iterations to converge, while learning-based approaches have limited generalization beyond the training distribution.

\subsection{LLM-Aided HLS Design}

LLM-aided design (LAD) has garnered significant attention in low-level HDL code generation and verification \cite{lik,llm_golden}, which has also catalyzed research interest in HLS. Previous work has incorporated retrieval-augmented generation (RAG) to provide HLS domain knowledge, with the aim of debugging and design optimization \cite{hlsdebugger, HLSPilot, ralad, iDSE}. However, RAG struggles to provide accurate search results, which may impair the reasoning of LLMs due to partially matched contexts \cite{icrag, gnn_rag}. 
Therefore, some efforts have improved the accuracy of generating HLS-C from natural language by fine-tuning LLMs. While these approaches support syntax and function error correction, they lack a comprehensive analysis of HLS compatibility issues.

Some studies treated LLMs as directive comparators within a Bayesian optimization framework \cite{directive_opt}, or adopted graph-level supervision to fine-tune LLMs \cite{LIFT} for HLS design optimization. While these methods demonstrate improvements in both optimization quality and efficiency, they fail to establish the interrelationship among HLS designs, directives, and corresponding QoR, limiting their effectiveness in design-specific optimization.

\section{Design \& Philosophy}
\label{sec:Methodology}
\begin{figure*}[t]
	\centering 
 	\includegraphics[width=16cm]{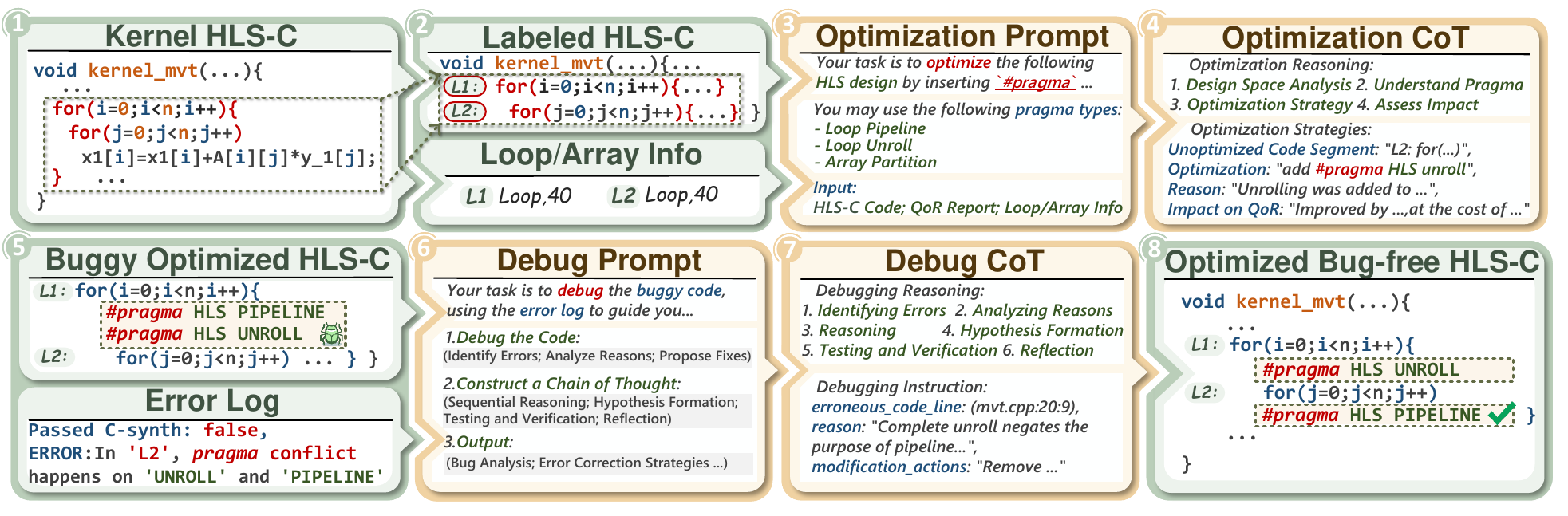} 
	\caption{An example of HLS-C optimization and error diagnosis in ChatHLS workflow.}
	\label{example}
    \vspace{-5pt}
\end{figure*}

\subsection{ChatHLS Architecture \& Workflow}

To address the aforementioned challenges, we propose the ChatHLS workflow to optimize HLS designs while incorporating robust code error correction capability. As illustrated in Figure \ref{workflow}.A, the architecture comprises two primary phases: \textit{HLS-C generation} and \textit{HLS-C debugging}.

In the \textbf{HLS-C generation} phase, LLM \bigcircled{1} leverages retrieved HLS-related context to transform input C algorithms or natural language descriptions.
A fine-tuned LLM \bigcircled{2} then selects effective HLS directive combination strategies and inserts directives within the specific structure (e.g., loops and arrays) of the generated HLS-C.
However, the inherent hallucinations of LLMs and their misalignment with HLS specifications may introduce errors during optimization, such as invalid pointer usage and type confusion during directive insertion.

The \textbf{HLS-C debugging} phase is designed to ensure the correctness of the generated HLS-C within ChatHLS workflow. Initially, the generated code is tested by the HLS tool. Upon detection of errors during C simulation and synthesis, we parse the compilation report and pair it with the erroneous code for a fine-tuned LLM \bigcircled{3} specifically tailored for error diagnosis. This model formulates explicit modification instructions with detailed analysis, which are then passed to LLM \bigcircled{4}. Operating under strict instruction adherence, this agent adopts the instructions to implement debugging.

For errors beyond the training distribution, we forward the parsed error message to LLM Group \bigcircled{5} for multifaceted evaluation. Subsequently, LLM \bigcircled{6} evaluates the proposed solutions and selects the most appropriate one to repair the code. Furthermore, the errors encountered at this stage are collected via a self-evolving framework. This allows ChatHLS to tackle a broader spectrum of complex HLS errors beyond generic C++ debugging.

\subsection{HLSTuner}

As a core component of the HLS-C generation, we propose HLSTuner to automate HLS directive selection, combination, configuration and insertion. To navigate the expansive design space of HLS-C, HLSTuner enables \textit{QoR-aware reasoning} to align optimization goals with hardware constraints.

\textbf{HLSTuner Architecture.} The input pair to HLSTuner includes source HLS-C, design metadata (e.g., array dimensions and loop trip counts), and initial QoR, including latency (cycles) and resource utilization percentages (DSP, LUT, FF), as shown in Figure \ref{example}. HLSTuner processes these inputs to select appropriate directives that match specific HLS design structures. It then coarsely estimates their impact on resource consumption, such as: \textit{"apply \texttt{PIPELINE} for deeply nested loops or inner loops with high trip counts significantly increases DSP and LUT resource consumption."} Following in-context guidance, HLSTuner formulates a detailed plan that specifies: (1) the combination of HLS directives with their types and factors, (2) the target code segments for optimization, and (3) the insertion actions. Finally, an insertion agent executes this plan for HLS-C optimization.

When the initial attempt fails to meet the desired performance, HLSTuner activates an iterative refinement incorporating current directives and the resulting QoR. HLSTuner analyzes QoR to scale loop parallelism up or down. For example, if hardware utilization exceeds the budget, HLSTuner halves parallelism and modifies related directives to support memory accesses. Conversely, if significant idle DSP or LUT resources exist, it prioritizes increasing the parallelism of deeply nested loops.

\begin{table*}[t]
\centering
\begin{adjustbox}{width=\textwidth}
\renewcommand{\arraystretch}{1.57}
\fontsize{9pt}{10pt}\selectfont
\setlength{\tabcolsep}{1mm}
\begin{tabular}
{
>{\centering\arraybackslash}m{1.68cm}
>{\centering\arraybackslash}m{2.4cm}
>{\raggedright\arraybackslash}m{4.2cm}
>{\raggedright\arraybackslash}p{10.5cm}
}
\hline\hline 
\textbf{Category} & \textbf{Error Type} & \multicolumn{1}{c}{\textbf{Error Message}} & \multicolumn{1}{c}{\textbf{Debugging Instruction}} \\
\hline 

\multirow{4}{1.6cm}[-1.2mm]{\centering HLS-C\\ Incompatible Errors} &
Dynamic Array Allocation (DAA) & \textbf{Error:} In function \texttt{A}: Undefined function \texttt{malloc} &  
\parbox[l]{10.5cm}{\textbf{Cause:} Dynamic memory allocation is not synthesizable. $\bm{\Rightarrow}$ \textbf{Diagnosis:} Replace dynamic allocation \texttt{malloc()} with fixed-size static array \texttt{A[]}.} \\   \cline{2-4} 

& Loop Index Out of Bounds (OOB) & \textbf{Error:} C TB testing failed, stop generating test vectors & 
\parbox[l]{10.5cm}{\textbf{Cause:} Out-of-bounds access creates faulty hardware, failing HLS co-simulation. $\bm{\Rightarrow}$ \textbf{Diagnosis:} Analyze array access patterns and correct loop boundary \texttt{<=} to \texttt{<}.} \\  \cline{2-4} 

& Pointer Access Error (PTR) & \textbf{Error:} @E Simulation failed: SIGSEGV & 
\parbox[l]{10.5cm}{\textbf{Cause:} Unconstrained pointers are not synthesizable. $\bm{\Rightarrow}$ \textbf{Diagnosis:} Replace unsafe pointer \texttt{*p} with explicit static array \texttt{p[]} to produce determined hardware.} \\ 

\hline 

\multirow{4}{1.6cm}[-1.3mm]{\centering HLS Directive Errors} &
Dataflow-Pipeline Conflict (DPC) & \textbf{Error:} \texttt{PIPELINE} and \texttt{DATAFLOW} are incompatible & 
\parbox[l]{10.5cm}{\textbf{Cause:} Apply conflict directives at same scope. $\bm{\Rightarrow}$ \textbf{Diagnosis:} Resolve producer-consumer dependency by removing \texttt{DATAFLOW} from logically interdependent loop.} \\ \cline{2-4} 

& Multi-Layer Pipeline (MLP) & \textbf{Error:} Forced nested loop full \texttt{UNROLL} cause synth time-out &
\parbox[l]{10.5cm}{\textbf{Cause:} \texttt{PIPELINE} on deep nested or large footprint loops cause resource explosion. $\bm{\Rightarrow}$ \textbf{Diagnosis:} Analyze loop structure and restrict \texttt{PIPELINE} to critical inner loops.} \\  \cline{2-4} 

& Array Partition Invalid Dim (AID) & \textbf{Warning:} \texttt{PARTITION} failed: size mismatch or dim too deep & 
\parbox[l]{10.5cm}{\textbf{Cause:} \texttt{PARTITION} exceeds declared dimensions. $\bm{\Rightarrow}$ \textbf{Diagnosis:} Correct dim parameter to match array declaration and intended memory access pattern.} \\

\hline\hline
\end{tabular}
\end{adjustbox}

\caption{Examples of \textit{BugRAG} entries and representative HLS-specific error types.}
\label{hls_errors}
\vspace{-10pt}
\end{table*}

\textbf{Training Strategy.}
The parallelism between loop execution and memory access primarily determines the HLS design performance. Therefore, our training aims to enable LLMs to understand the semantics of HLS directives such as \texttt{PIPELINE}, \texttt{UNROLL}, and \texttt{ARRAY\_PARTITION} to optimize these structures, including their usage, effects, and interplay. To enable the LLM to capture the causal relationship between directive changes and QoR variations, we move beyond simply mapping the source code to the optimized code by instilling a \textit{QoR-aware reasoning} capability. This reasoning explicitly provides a step-by-step rationales linking directive modifications to synthesized hardware architecture and performance gains. We use optimization chain-of-thought (CoT) generated by a teacher model to supervise this training.

As shown in Figure \ref{workflow}.B, we employ NSGA-\uppercase\expandafter{\romannumeral2} to generate diverse optimized HLS designs and collect the corresponding QoR reports \cite{CollectiveHLS}. These samples are structured for LLM \bigcircled{7} to generate the optimization CoT. This CoT analyzes the QoR variations (changes in latency and resource usage). It identifies data dependencies to justify \texttt{PIPELINE} choices or balance memory bandwidth with \texttt{ARRAY\_PARTITION} and evaluate hardware parallel processing architectures. Finally, we construct a dataset that pairs source HLS-C, inserted directives, and the CoT. This equips LLM \bigcircled{2} with \textit{QoR-aware reasoning}, enabling rapid identification of high-performance optimization strategies within specific resource constraints.

\subsection{HLSFixer}

At the core of the HLS-C debugging phase, we develop HLSFixer, a hierarchical code repair framework designed to address syntax incompatibility and directive misuse errors during HLS-C generation and optimization. Within this framework, an analysis agent adopts a \textit{reasoning-to-instruction} method, analyzing the HLS tool feedback to formulate error modification actions. For errors beyond the scope of the single analysis model, HLSFixer supports an LLM-as-a-judge system that augments the instruction to improve debugging accuracy.

\textbf{HLSFixer Architecture.}
As depicted in Figure \ref{example}, debugging begins with HLS simulation and synthesis log parsing, which uses keyword matching to extract error messages. The analysis LLM \bigcircled{3} then examines the error causes and provides debugging instructions, as illustrated in Table \ref{hls_errors}. After applying targeted modifications to the HLS-C based on this analysis, HLSFixer retests the corrected HLS design against the \textit{golden results} to ensure semantic equivalence with the original design intent.

When LLM \bigcircled{3} fails to correct errors, we implement an LLM-as-a-judge system to refine the debugging instructions. We provide the modified code and error messages to LLM Group \bigcircled{5}, which generates diverse debugging instructions. These candidates are then compiled and evaluated by the scoring agent \bigcircled{6} (functioning as the judge). Based on clarity, logical soundness, alignment with error messages, and the scope of code modification, this agent selects the optimal suggestion to improve the quality of debugging instruction feedback.

\textbf{Training Strategy.}
To train the analysis agent (LLM \bigcircled{3}) to perform reasoning and generate accurate debug instructions, we decouple debugging into error identification, diagnosis and repair, as shown in Figure \ref{workflow}.C 
Specifically, we construct testbench with verified \textit{golden results} to extract key error messages. These messages enable the LLM to pinpoint error locations and propose targeted modifications. Subsequently, we prompt LLM \bigcircled{9} to review the correct HLS-C code, pairing buggy code segments with corresponding error messages to construct debugging CoT. The generated CoT analyzes within \texttt{<reasoning>} how the specific error types lead to the error message, and reflects on the code after assuming the fix to ensure the modification is reasonable. The \texttt{<instruction>} pinpoints the exact error line, explains the cause, and details the precise repair action. Finally, we construct a dataset comprising buggy code, associated error messages, and generated CoT. 

We further employ direct preference optimization (DPO) \cite{DPO}, which learns implicit reward functions from a preference dataset composed of CoT and non-CoT pairs. For rejected non-CoT, we remove error messages when LLM generates the debug CoT. This may confuse the relationship between HLS-specific errors and error messages. This method enables the fine-tuned LLM \bigcircled{3} to rely more closely on parsed error messages during debugging, thereby aligning with expert error analysis preferences.

\begin{figure}[t]
	\centering 
\includegraphics[width=7.8cm]{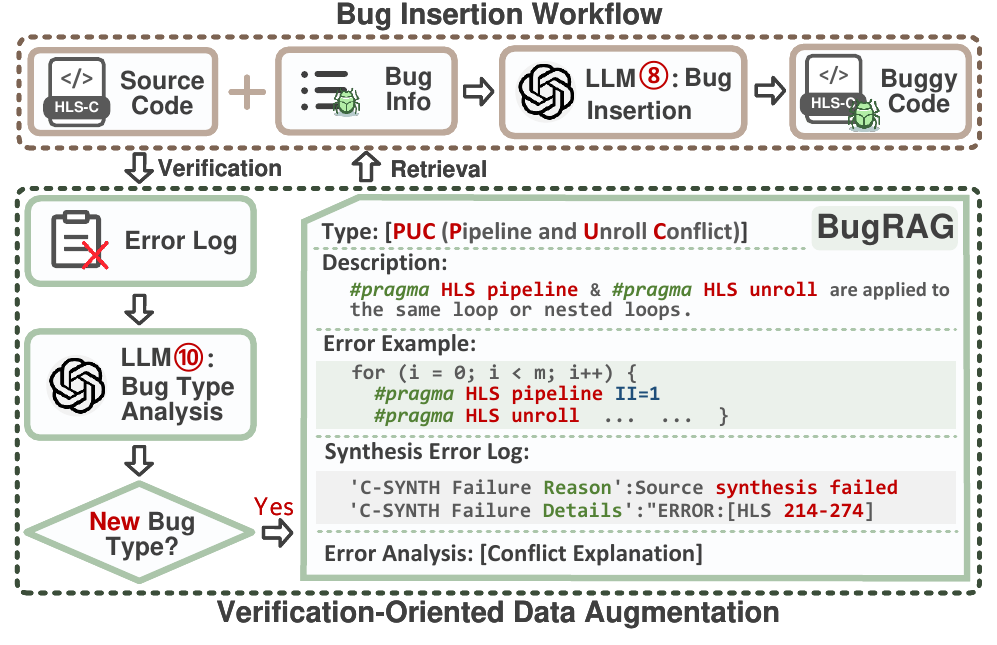} 
	\caption{Verification dataset construction workflow.}
	\label{RAG}
    \vspace{-13pt}
\end{figure}

\subsection{Verification-Oriented Data Augmentation}

Given that LLMs struggle with strict hardware constraints of synthesizable HLS-C and HLS dataset scarcity, we propose the \textit{Verification-Oriented Data Augmentation (VODA)} paradigm. VODA leverages HLS tool feedback to enable the LLM to learn from failures and self-evolve to address more complex errors in the debugging phase. The philosophy of VODA is to construct and progressively expand a repository of buggy code, featuring various syntax and logic errors. Each entry in the repository is annotated with its error message and detailed root-cause analysis.

\textbf{BugRAG.}
We design \textit{BugRAG} that dynamically collects and expands the range of HLS-specific error types. Its construction is based on a comprehensive analysis of AMD forum inquiries, prior research \cite{HeteroGen, automated_c++, Chrysalis-HLS}, and VODA-expanded error cases. We categorize errors encountered during HLS-C generation and optimization phases, as representative cases shown in Table \ref{hls_errors}. 
We define \textit{HLS-C Incompatible Errors} as syntax (e.g., dynamic arrays and out-of-bounds array access) that, while valid in standard C, result in incorrect hardware behavior after synthesis. We classify \textit{HLS Directive Errors} as incorrect placement, combination, or configuration of HLS directives, leading to invalid optimizations or synthesis failures.
These cases are structured into modular error slices within \textit{BugRAG}, incorporating mnemonic identifiers to improve retrieval accuracy \cite{Chrysalis-HLS}. We provide a list of collected error types in Appendix \ref{appendix:voda}.

\textbf{VODA Workflow.}
Based on \textit{BugRAG}, VODA operates in two stages. The first stage is the continuous expansion of error cases to populate the error repository, as illustrated in Figure \ref{RAG}. When an HLS design fails verification, an inspection agent (LLM \bigcircled{10}) examines the erroneous code and error messages parsed from the HLS tool test results. It then generates an error slice containing descriptions, examples, and analysis of a specific HLS error type and queries \textit{BugRAG} to check for existing entries. If unmatched, the inspection agent identifies a new error type and integrates the slice into the error repository with a new mnemonic identifier.

In the second stage, we generate a verification dataset through a controlled bug injection process. This process is facilitated by an insertion agent (LLM \bigcircled{8}), which generates buggy code by integrating retrieved error slices from \textit{BugRAG} as context. The agent assesses the contextual applicability of potential bugs, reducing the probability that the LLM forcibly generates trivial results.

\section{Evaluations}
\label{sec:Evaluation}
\subsection{Dataset Construction \& LLM Training}

For HLSFixer, we constructed 10,878 buggy code covering 33 error types to train LLM \bigcircled{3} through SFT and 3,716 preference pairs for DPO. These buggy samples were injected from 35 base designs from \textit{Kernel} \cite{polybenchC} and \textit{Vitis} \cite{vitis_examples} sources. For HLSTuner, we collected 4,804 samples from 20 Rosetta kernels \cite{CollectiveHLS} to fine-tune LLM \bigcircled{2}. Appendix \ref{appendix:similarity} details the train-test similarity analysis. All data were generated by DeepSeek-V3.2.

The training was conducted on 8$\times$ NVIDIA H800-80G GPUs, based on Qwen-2.5-Coder-14B-Instruct. We used full parameter tuning, AdamW optimizer, 3 epochs with 1e-5 learning rate for SFT, and 2 epochs with 5e-6 learning rate for DPO. Training details are provided in Appendix \ref{appendix:training_detail}.

\subsection{Benchmarks \& Metrics}

We evaluated ChatHLS on 108 natural language to HLS-C tasks from \cite{hlseval} and our own (e.g., FSMs, counters, rotate operators), each with a validation testbench. For HLSFixer, we constructed 591 test cases from \textit{Kernel}, \textit{Vitis}, and \textit{Manual} (15 manual designs).

For each test case $c_{i}$, we adopt $pass@k$ to quantify the HLS-C generation and debugging performance. It estimates the probability that at least one correct solution $c_{i}^*$ is found among $k$ generated samples. We set $n$ trials per $c_{i}$ ($n \geq k$). A trial is correct if it passes HLS toolchain verification, including C-Simulation (CSIM), Synthesis (CSYN), and C/RTL Co-simulation (COSIM). Here, CSIM runs the C/C++ testbench against the HLS-C code to validate functional correctness, CSYN compiles HLS-C into RTL and reports timing/resource QoR, and COSIM executes the same testbench against the RTL to verify functional equivalence between the high-level design and generated hardware. 
\begin{equation}
\label{eq:pass_at_k}
pass@k := \mathbb{E}_{i} \left[ 1 - \frac{\binom{n-c_{i}^*}{k}}{\binom{n}{k}} \right]
\end{equation}
We generate $n = 5$ candidates for each debugging case using $pass@1$. For HLS-C generation, we set $n = 20$ to evaluate the pass rate in stages.

We evaluated HLSTuner on various HLS kernels, including linear algebra, cryptographic algorithms, and neural network accelerators \cite{hlsfactory}. Synthesis was performed with Vitis HLS 2022.1 targeting the Xilinx ZCU106 MPSoC at 100 MHz. The QoR includes execution latency (cycles), utilization of digital signal processors (DSP), flip-flops (FF) and look-up tables (LUT). To quantify HLSTuner performance, we utilized Vitis HLS auto pipeline optimization as a baseline and measured $Speedup$, as defined in Appendix \ref{appendix:HLSTuner}.

In practice, each kernel requires two full verification passes (CSIM/CSYN/COSIM for generation and debugging) and up to five CSYN iterations for optimization. The runtime remains within practical budgets: $<3$ minutes for generation, $<10$ minutes for debugging, and $<30$ minutes for optimization.

\begin{figure}[t]
	\centering 
\includegraphics[width=7.8cm]{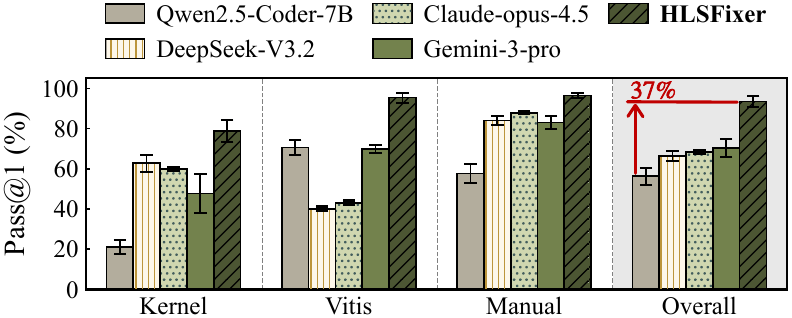} 
	\caption{Comparison of debugging capability between HLSFixer and general-purpose LLMs.}
	\label{fig:hlsfixer}
\end{figure}

\begin{table}[t]
  \centering
  \fontsize{9pt}{10pt}\selectfont
  \renewcommand{\arraystretch}{1.3}
  \begin{adjustbox}{width=0.47\textwidth}
  \setlength{\tabcolsep}{2.5pt} 
  \begin{tabular}{lcccccc}
\hline\hline 
& \multicolumn{2}{c}{CSIM} & \multicolumn{2}{c}{CSYN} & \multicolumn{2}{c}{COSIM} \\
\cmidrule(lr){2-3} \cmidrule(lr){4-5} \cmidrule(lr){6-7}
Model & Pass@1 & Pass@5 & Pass@1 & Pass@5 & Pass@1 & Pass@5 \\
\hline
DeepSeek-V3.2 & 47.0\% & 56.8\% & 43.2\% & 55.1\% & 31.5\% & 45.2\% \\
Gemini-3-pro & 57.9\% & 68.7\% & 56.5\% & 68.2\% & 48.1\% & 60.0\% \\
ChatHLS (w/RAG) & {59.0\%} & {77.7\%} & {57.3\%} & {75.4\%} & {52.3\%} & {70.3\%} \\
\cellcolor[HTML]{E2E7CF}ChatHLS (w/HLSFixer) & \cellcolor[HTML]{E2E7CF}\textbf{82.1\%} & \cellcolor[HTML]{E2E7CF}\textbf{90.1\%} & \cellcolor[HTML]{E2E7CF}\textbf{81.2\%} & \cellcolor[HTML]{E2E7CF}\textbf{90.0\%} & \cellcolor[HTML]{E2E7CF}\textbf{77.2\%} & \cellcolor[HTML]{E2E7CF}\textbf{87.6\%} \\
\hline\hline 
  \end{tabular}
  \end{adjustbox}
\caption{Comparison of the generation capability.}
  \label{table:generation}
  \vspace{-8pt}
\end{table}

\subsection{HLSFixer Capability Analysis}

\subsubsection{Comparison with General LLM.} 
We compare HLSFixer with general-purpose LLMs for error analysis. Figure \ref{fig:hlsfixer} shows that across comprehensive test cases, HLSFixer achieves a 93.4\% pass@1, outperforming Claude-opus-4.5 by 36.8\% and Gemini-3-pro by \textbf{32.6\%}. This gap highlights the specialized debugging reasoning of HLSFixer, which enables more accurate identification, analysis, and correction of HLS-specific errors. We provide a comparison of models with different parameter scales in the Appendix \ref{appendix:Scale}.

\begin{figure}[t]
	\centering 
\includegraphics[width=7.8cm]{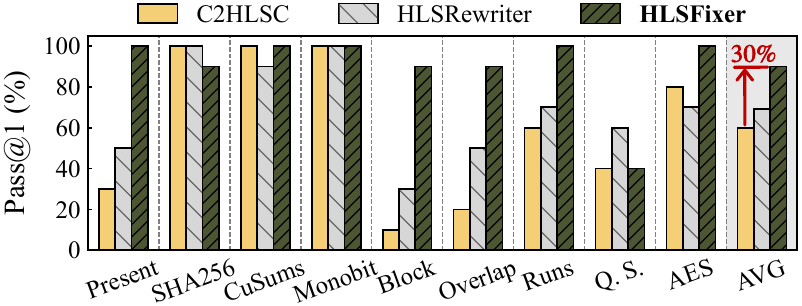} 
	\caption{Comparison of HLS-C generation pass rate.}
	\label{c2hlsc}
    \vspace{-8pt}
\end{figure}

With HLSFixer, ChatHLS shows a remarkable improvement in pass rates for HLS-C generation tasks, as shown in Table \ref{table:generation}. By incorporating RAG (using DeepSeek-V3.2 as LLM \bigcircled{1}) to provide basic HLS programming specification, combined with a multi-agent debugging approach, ChatHLS ensures the generation robustness, outperforming single Gemini-3-pro by \textbf{41.8\%}. 
As shown in Figure \ref{c2hlsc}, we compare ChatHLS with C2HLSC \cite{C2HLSC} and HLSRewriter \cite{hlsrewriter}. These results highlight the efficacy of our framework in navigating the complexities of HLS constraints. Appendix \ref{app:rag_fairness} highlights the debugging advantages of HLSFixer in internalizing HLS-specific reasoning beyond retrieval alone.

\begin{figure}[t]
	\centering 
	\includegraphics[width=7.8cm]{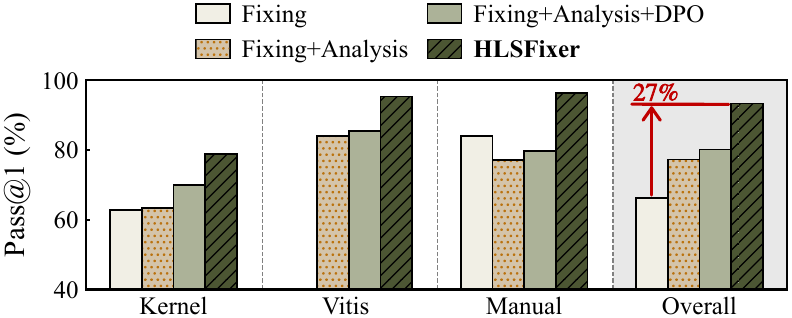} 
	\caption{Ablation study of HLSFixer design.}
	\label{Ablation}
\end{figure}

\begin{figure*}[!t]
	\centering 
\includegraphics[width=\textwidth]{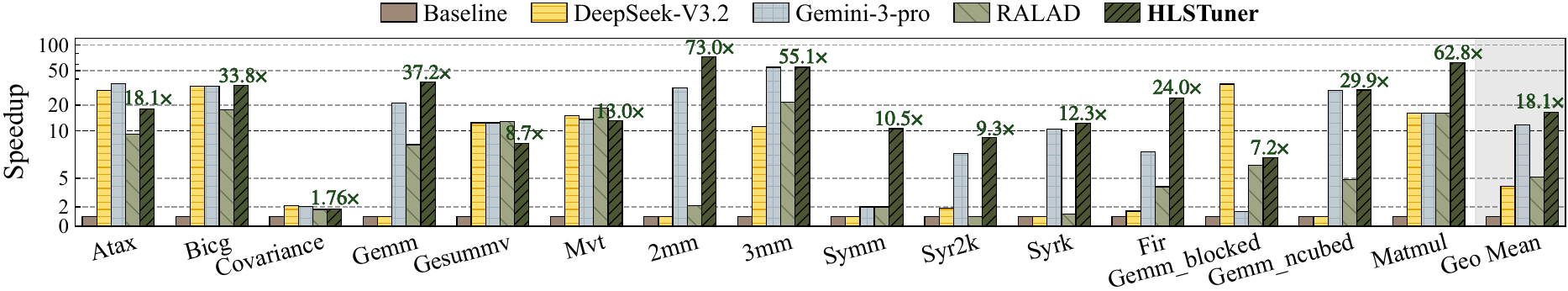} 
	\caption{Comparison of optimization capability between Vitis HLS auto optimization (Baseline), general-purpose LLMs, retrieval-augmented method (RALAD) and HLSTuner.}
	\label{fig:performance}
\end{figure*}

\subsubsection{Ablation Study.}
We compared HLSFixer with (1) fixing only with DeepSeek-V3.2, (2) fixing combined with analysis agent using the fine-tuned LLM \bigcircled{3}, and (3) fixing and analysis agent augmented with DPO, as shown in Figure \ref{Ablation}. After applying fine-tuned analysis agent using \textit{reasoning-to-instruction}, the repair pass rate increased by 16.6\% compared to using a single fixing agent.
Augmenting LLM \bigcircled{3} with DPO further increased the overall pass rate by 3.7\%. For errors unresolved in a single attempt, we performed multifaceted evaluations, which led to an additional 16.5\% improvement. 
Three debugging instructions from different LLMs (GPT-5, Claude-opus-4.5, and Qwen3-8B in this experiment) were evaluated by a long-context LLM (Gemini-3-Pro). These results validate the effectiveness of the fine-tuned analysis model and multi-LLM verification system for HLS-specific error debugging.

\begin{table}[t]
\centering
\fontsize{9pt}{10pt}\selectfont
\renewcommand{\arraystretch}{1.2}
\begin{adjustbox}{width=0.47\textwidth}
\setlength{\tabcolsep}{4pt} 
\begin{tabular}{llrrr}
\hline\hline
\textbf{Accelerators} & \textbf{Metric} & \cellcolor[HTML]{E2E7CF}\textbf{HLSTuner} & \textbf{RALAD} & \textbf{Baseline} \\ 
\hline
\multirow{5}{*}{\textbf{MobileNet}} 
 & Latency & \cellcolor[HTML]{E2E7CF}\textbf{5.88M}& 13.33M & 13.14M   \\
 & DSP& \cellcolor[HTML]{E2E7CF}573 (33.2\%) & 729 (42.2\%) & 69 (4.0\%)  \\
 & FF & \cellcolor[HTML]{E2E7CF}36.6K (8.0\%)   & 108.7K (23.6\%) & 4.5K (1.0\%)\\
 & LUT & \cellcolor[HTML]{E2E7CF}69.5K (30.1\%)   & 111.1K (48.2\%) & 12.8K (5.6\%) \\
 & Speedup& \cellcolor[HTML]{E2E7CF}\textbf{2.233$\times$} & 0.986$\times$ & 1.000$\bm\times$ \\ 
 \hline
\multirow{5}{*}{\textbf{Transformer}} 
 & Latency & \cellcolor[HTML]{E2E7CF}\textbf{68.51K} & 88.34K & 83.30K   \\
 & DSP& \cellcolor[HTML]{E2E7CF}1215 (70.3\%) & 274 (15.9\%)  & 222 (12.8\%)  \\
 & FF & \cellcolor[HTML]{E2E7CF}207.1K (44.9\%)   & 108.3K (23.5\%) & 95.1K (20.6\%) \\
 & LUT & \cellcolor[HTML]{E2E7CF}165.0K (71.6\%)   & 133.3K (57.8\%) & 84.0K (36.4\%) \\
 & Speedup& \cellcolor[HTML]{E2E7CF}\textbf{1.216$\bm\times$} & 0.943$\times$ & 1.000$\times$ \\ 
 \hline\hline
\end{tabular}
\end{adjustbox}
\caption{Comparison of performance optimization results in hardware accelerators. HLSTuner achieves significant speedup over baseline and RAG-based method.}
\label{table:complex}
\end{table}

\begin{table}[t]
\centering
\small
\fontsize{9pt}{10pt}\selectfont
\setlength{\tabcolsep}{13pt}
\renewcommand{\arraystretch}{1.2}
\resizebox{\columnwidth}{!}{
\begin{tabular}{lccc}
\hline
\hline
Post-Implementation Metric & Min & Mean & Max \\
\hline
Critical path (ns) & 2.768 & 5.859 & 8.328 \\
Total on-chip power (W) & 0.598 & 0.705 & 0.862 \\
\hline
\hline
\end{tabular}}
\caption{Implementation validity of the best HLSTuner designs across 16 workloads. All designs satisfy the target frequency (critical path $<10$\,ns) and remain within the acceptable on-chip power budget ($<0.9$\,W).}
\label{tab:impl_validity}
\end{table}

\subsection{HLSTuner Capability Analysis}

\subsubsection{Comparison with General LLMs.}
Figure \ref{fig:performance} compares HLSTuner with Gemini-3-pro, DeepSeek-V3.2, and RALAD (reproduced using GPT-5.1) \cite{ralad} in achieving the optimal optimization of HLS kernel performance within 15 trials.
HLSTuner achieved a geometric mean speedup of 18.1$\times$ compared to the baseline, 4.0$\times$ over DeepSeek-V3.2, 1.5$\times$ over Gemini-3-pro and 3.3$\times$ over RALAD, while maintaining resource utilization \textit{under 80\%} on target hardware. Current general-purpose LLMs often produce excessive resource utilization or even synthesis failures. In contrast, HLSTuner, driven by a fine-tuned agent with specialized hardware optimization knowledge, achieves optimal speedups by intelligently tuning directive combinations and configurations. Furthermore, HLSTuner deepens its understanding of directive semantics, minimizing errors that may arise during directive insertion. We provide a detailed analysis in Appendix \ref{appendix:hlstuner}.

\begin{figure}[t]
	\centering 
\includegraphics[width=7.8cm]{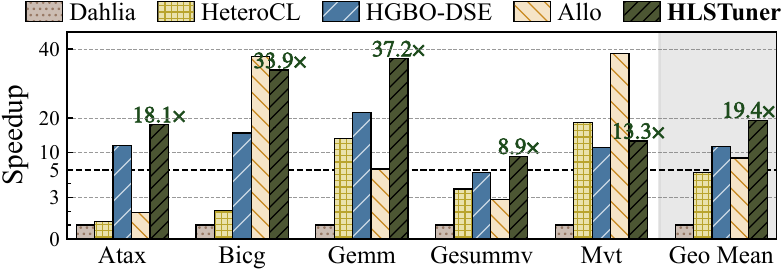} 
	\caption{Latency speedup of DSL-based (Dahlia, HeteroCL, Allo), learning-based (HGBO-DSE) methods.}
	\label{DSE2}
\end{figure}

To assess HLSTuner on more complex designs, we applied it to MobileNet and Transformer. Table \ref{table:complex} shows that HLSTuner obtains notable speedups over Vitis HLS baseline. Compared to RALAD, it delivers a \textbf{2.3$\bm\times$} speedup for MobileNet, and a 1.3$\times$ speedup for Transformer. This highlights the benefit of \textit{QoR-aware reasoning}, which enables LLMs to control optimization trajectories by iteratively sensing the impact of directives on QoR, thereby improving performance while meeting acceptable hardware resource constraints. Furthermore, we observed that even with GPT-5.1, RALAD struggles to comprehend the usage rules and interactions between directives. Out of 15 trials, most resulted in inefficient optimization due to directive misuse or mismatched loop unrolling and array partitioning. 

ChatHLS does not invoke algorithm-level C/C++ refactoring during optimization. In HLS, unconstrained structural rewriting may introduce functional deviations that are difficult to attribute from synthesis logs alone. By restricting HLSTuner to directive-level transformations, ChatHLS keeps the kernel semantics unchanged while still covering the standard HLS design space exploration problem of pragma selection, insertion, and factor tuning. Table \ref{tab:impl_validity} reports post-implementation results after Place-and-Route. All optimized designs satisfy the target frequency, showing that the reported latency reduction in cycles is physically realizable rather than an artifact of infeasible unrolling. Moreover, ChatHLS generates \textit{dedicated circuits}. This explains the workload-dependent resource variation in Table \ref{table:complex}: MobileNet and Transformer exhibit different arithmetic intensity and memory-access structures, and HLSTuner specializes pragma decisions to each design while keeping resource utilization within the target budget.

\subsubsection{Comparison with DSL-based and learning-based methods.}
Figure \ref{DSE2} compares HLSTuner against DSL-based \cite{Dahlia, HeteroCL, allo} and learning-based \cite{hgbo_dse} methods on the same HLS kernels. HLSTuner achieved a geometric mean speedup of 19.4$\times$ over Dahlia, 4.0$\times$ over HeteroCL, 2.3$\times$ over Allo and 1.6$\times$ over HGBO-DSE. Among them, DSL-based methods require users to invest considerable time in mastering the usage of primitives for optimization. Meanwhile, heuristic or learning-based DSE methods necessitate numerous iterations (100 searches for HGBO-DSE) to converge on an optimal solution. In contrast, HLSTuner bridges the gap in directive semantics comprehension between existing LLMs and HLS design optimization. This allows it to avoid evaluating ineffective directive combinations and configurations and rapidly identify promising optimization strategies. Moreover, this approach optimizes performance by exploring \textit{directive} parallelism without altering \textit{code} functional semantics, thus avoiding the correctness concerns associated with LLM-based code rewriting.

\section{Conclusion}
\label{sec:Conclusio}
This paper proposes ChatHLS, which features a reasoning-augmented error analysis model and an optimization model proficient in HLS directives to automate HLS-C generation and optimization. We strengthen the robustness of the system by continually expanding error cases through VODA.
Experimental results indicate that ChatHLS boosts HLS-C debugging pass rate by \textbf{32.6\%} relative to Gemini-3-pro. It also achieves a \textbf{3.3$\bm\times$} geo mean speedup against RAG-based methods across various HLS designs. These improvements pave the way for a more efficient and reliable hardware design.

\section*{Acknowledgement}

This work is supported by the National Natural Science Foundation of China (Grant No.92464301), the National Key Research and
Development Program (Grant No.2024YFB4405600), and the Key
Research and Development Program of Jiangsu Province (Grant
No.BG2024010).

\section*{Limitation}

1) As loop optimization constitutes the core performance bottleneck in scientific computing and neural network acceleration, ChatHLS focuses on tuning directives for loop execution parallelism. It lacks support for complex \texttt{DATAFLOW} control directives with producer-consumer logic or for inserting AXI interface directives to implement the HLS-C design as a valid IP for FPGA deployment.

2) While ChatHLS adjusts its optimization strategies during iteration under specific resource utilization constraints, its effectiveness has so far been validated on the Xilinx ZCU106 MPSoC. Broader case studies on FPGAs with different resource constraints would benefit its portability.

\section*{Ethical Considerations}

The LLMs fine-tuned and utilized are intended exclusively for the scientific purpose of HLS-C generation, debugging and optimization. We emphasize that LLM-generated HLS-C should undergo standard simulation and verification before deployment to ensure hardware safety. We confirm that both the fine-tuning and test datasets do not contain offensive or proprietary content. 

\bibliography{custom}

@inproceedings{automated_c++,
  title={{Automated C/C++ Program Repair for High-Level Synthesis via Large Language Models}},
  author={Xu, Kangwei and Zhang, Grace Li and Yin, Xunzhao and Zhuo, Cheng and Schlichtmann, Ulf and Li, Bing},
  booktitle={Proceedings of the 2024 ACM/IEEE International Symposium on Machine Learning for CAD},
  pages={1--9},
  year={2024}
}

@inproceedings{ChatCPU,
author = {Wang, Xi and Wan, Gwok-Waa and Wong, Sam-Zaak and Zhang, Layton and Liu, Tianyang and Tian, Qi and Ye, Jianmin},
title = {{ChatCPU: An Agile CPU Design and Verification Platform with LLM}},
year = {2024},
booktitle = {Proceedings of the 61st ACM/IEEE Design Automation Conference},
numpages = {6},
series = {DAC '24}
}

@article{C2HLSC,
author = {Collini, Luca and Garg, Siddharth and Karri, Ramesh},
title = {{C2HLSC: Leveraging Large Language Models to Bridge the Software-to-Hardware Design Gap}},
year = {2025},
volume = {30},
number = {6},
journal = {ACM Transactions on Design Automation of Electronic Systems},
articleno = {96},
numpages = {24}
}

@INPROCEEDINGS{HLSPilot,
  author={Xiong, Chenwei and Liu, Cheng and Li, Huawei and Li, Xiaowei},
  booktitle={2024 ACM/IEEE International Conference On Computer Aided Design (ICCAD)}, 
  title={{HLSPilot: LLM-Based High-Level Synthesis}}, 
  year={2024},
  volume={},
  number={},
  pages={1-9},
}

@INPROCEEDINGS{ralad,
  author={Xu, Haocheng and Hu, Haotian and Huang, Sitao},
  booktitle={2024 IEEE LLM Aided Design Workshop (LAD)}, 
  title={{Optimizing High-Level Synthesis Designs with Retrieval-Augmented Large Language Models}}, 
  year={2024},
  volume={},
  number={},
  pages={1-5},
}

@inproceedings{Dahlia,
author = {Nigam, Rachit and Atapattu, Sachille and Thomas, Samuel and Li, Zhijing and Bauer, Theodore and Ye, Yuwei and Koti, Apurva and Sampson, Adrian and Zhang, Zhiru},
title = {Predictable accelerator design with time-sensitive affine types},
year = {2020},
booktitle = {Proceedings of the 41st ACM SIGPLAN Conference on Programming Language Design and Implementation},
pages = {393–407},
numpages = {15}
}

@article{allo,
author = {Chen, Hongzheng and Zhang, Niansong and Xiang, Shaojie and Zeng, Zhichen and Dai, Mengjia and Zhang, Zhiru},
title = {{Allo: A Programming Model for Composable Accelerator Design}},
year = {2024},
issue_date = {June 2024},
volume = {8},
journal = {Proceedings of the ACM on Programming Languages},
articleno = {171},
numpages = {28},
}

@inproceedings{HeteroGen,
author = {Zhang, Qian and Wang, Jiyuan and Xu, Guoqing Harry and Kim, Miryung},
title = {{HeteroGen: transpiling C to heterogeneous HLS code with automated test generation and program repair}},
year = {2022},
booktitle = {Proceedings of the 27th ACM International Conference on Architectural Support for Programming Languages and Operating Systems},
pages = {1017–1029},
numpages = {13},
series = {ASPLOS '22}
}

@article{CollectiveHLS,
author = {Ferikoglou, Aggelos and Kakolyris, Andreas and Masouros, Dimosthenis and Soudris, Dimitrios and Xydis, Sotirios},
title = {{CollectiveHLS: A Collaborative Approach to High-Level Synthesis Design Optimization}},
year = {2024},
volume = {18},
number = {1},
journal = {ACM Transactions on Reconfigurable Technology and Systems},
articleno = {11},
numpages = {32}
}

@article{FPGAHLSToday,
author = {Cong, Jason and Lau, Jason and Liu, Gai and Neuendorffer, Stephen and Pan, Peichen and Vissers, Kees and Zhang, Zhiru},
title = {{FPGA HLS Today: Successes, Challenges, and Opportunities}},
year = {2022},
issue_date = {December 2022},
volume = {15},
number = {4},
journal = {ACM Transactions on Reconfigurable Technology and Systems},
articleno = {51},
numpages = {42},
}

@misc{polybenchC,
  title={{PolyBench/C 4.2}},
  author={Pouchet, Louis-Noel and Yuki, Tomofumi},
  year = {2016},
  url = {http://polybench.sf.net},
  lastchecked = {2024-11-16}
}

@misc{vitis_examples,
    title={{Vitis-HLS-Introductory-Examples}},
    author = {{Xilinx Inc.}},
    url = {https://github.com/Xilinx/Vitis-HLS-Introductory-Examples},
    year = {2024}
}

@inproceedings{DPO,
author = {Rafailov, Rafael and Sharma, Archit and Mitchell, Eric and Ermon, Stefano and Manning, Christopher D. and Finn, Chelsea},
title = {Direct preference optimization: your language model is secretly a reward model},
year = {2024},
booktitle = {Proceedings of the 37th International Conference on Neural Information Processing Systems}
}

@INPROCEEDINGS{heterorefactor,
  author={Lau, Jason and Sivaraman, Aishwarya and Zhang, Qian and Gulzar, Muhammad Ali and Cong, Jason and Kim, Miryung},
  booktitle={2020 IEEE/ACM 42nd International Conference on Software Engineering (ICSE)}, 
  title={{HeteroRefactor: Refactoring for Heterogeneous Computing with FPGA}}, 
  year={2020},
  volume={},
  number={},
  pages={493-505},
  doi={}
  }

@ARTICLE{DSEReview,
  author={Schafer, Benjamin Carrion and Wang, Zi},
  journal={IEEE Transactions on Computer-Aided Design of Integrated Circuits and Systems}, 
  title={High-Level Synthesis Design Space Exploration: Past, Present, and Future}, 
  year={2020},
  volume={39},
  number={10},
  pages={2628-2639},
}

@INPROCEEDINGS{Chrysalis-HLS,
  author={Wan, Lily Jiaxin and Huang, Yingbing and Li, Yuhong and Ye, Hanchen and Wang, Jinghua and Zhang, Xiaofan and Chen, Deming},
  booktitle={2024 29th Asia and South Pacific Design Automation Conference (ASP-DAC)}, 
  title={Software/Hardware Co-design for {LLM} and Its Application for Design Verification}, 
  year={2024},
  volume={},
  number={},
  pages={435-441},
}

@article{AutoDSE,
author = {Sohrabizadeh, Atefeh and Yu, Cody Hao and Gao, Min and Cong, Jason},
title = {{AutoDSE: Enabling Software Programmers to Design Efficient FPGA Accelerators}},
year = {2022},
volume = {27},
number = {4},
journal = {ACM Transactions on Design Automation of Electronic Systems},
articleno = {32},
numpages = {27}
}

@inproceedings{HeteroCL,
author = {Lai, Yi-Hsiang and Chi, Yuze and Hu, Yuwei and Wang, Jie and Yu, Cody Hao and Zhou, Yuan and Cong, Jason and Zhang, Zhiru},
title = {{HeteroCL: A Multi-Paradigm Programming Infrastructure for Software-Defined Reconfigurable Computing}},
year = {2019},
booktitle = {Proceedings of the 2019 ACM/SIGDA International Symposium on Field-Programmable Gate Arrays},
pages = {242–251},
numpages = {10}
}

@INPROCEEDINGS{scalehls,
  author={Ye, Hanchen and Hao, Cong and Cheng, Jianyi and Jeong, Hyunmin and Huang, Jack and Neuendorffer, Stephen and Chen, Deming},
  booktitle={2022 IEEE International Symposium on High-Performance Computer Architecture (HPCA)}, 
  title={{ScaleHLS: A New Scalable High-Level Synthesis Framework on Multi-Level Intermediate Representation}}, 
  year={2022},
  volume={},
  number={},
  pages={741-755}
}

@inproceedings{moe_dse,
  title={Hierarchical mixture of experts: Generalizable learning for high-level synthesis},
  author={Li, Weikai and Wang, Ding and Ding, Zijian and Sohrabizadeh, Atefeh and Qin, Zongyue and Cong, Jason and Sun, Yizhou},
  booktitle={Proceedings of the AAAI Conference on Artificial Intelligence},
  pages={18476--18484},
  year={2025}
}

@inproceedings{hgbo_dse,
  title={Hgbo-dse: Hierarchical gnn and bayesian optimization based hls design space exploration},
  author={Kuang, Huizhen and Cao, Xianfeng and Li, Jingyuan and Wang, Lingli},
  booktitle={2023 International Conference on Field Programmable Technology (ICFPT)},
  pages={106--114},
  year={2023}
}

@article{chipmind,
title={{ChipMind: Retrieval-Augmented Reasoning for Long-Context Circuit Design Specifications}}, 
volume={40}, 
number={2}, 
journal={Proceedings of the AAAI Conference on Artificial Intelligence}, 
author={Xing, Changwen and Wong, SamZaak and Wan, Xinlai and Lu, Yanfeng and Zhang, Mengli and Ma, Zebin and Qi, Lei and Li, Zhengxiong and Guan, Nan and Jiang, Zhe and Wang, Xi and Yang, Jun}, 
year={2026}, 
pages={1337-1345} 
}

@INPROCEEDINGS{hlseval,
  author={Abi-Karam, Stefan and Hao, Cong},
  booktitle={2025 IEEE International Conference on LLM-Aided Design (ICLAD)}, 
  title={{HLS-Eval: A Benchmark and Framework for Evaluating LLMs on High-Level Synthesis Design Tasks}}, 
  year={2025},
  volume={},
  number={},
  pages={219-226}
}

@article{hlsrewriter,
  title={{HLSRewriter: Efficient Refactoring and Optimization of C/C++ Code with LLMs for High-Level Synthesis}},
  author={Xu, Kangwei and Zhang, Grace Li and Yin, Xunzhao and Zhuo, Cheng and Schlichtmann, Ulf and Li, Bing},
  journal={ACM Transactions on Design Automation of Electronic Systems},
  year={2025}
}

@article{hlsdebugger,
  title={{HLSDebugger: Identification and Correction of Logic Bugs in HLS Code with LLM Solutions}},
  author={Wang, Jing and Liu, Shang and Lu, Yao and Xie, Zhiyao},
  journal={2025 IEEE/ACM International Conference on Computer Aided Design (ICCAD)},
  year={2025}
}

@inproceedings{lik,
  title={{Location is Key: Leveraging LLM for Functional Bug Localization in Verilog Design}},
  author={Yao, Bingkun and Wang, Ning and Zhou, Jie and Wang, Xi and Gao, Hong and Jiang, Zhe and Guan, Nan},
  booktitle={2025 62nd ACM/IEEE Design Automation Conference (DAC)},
  pages={1--7},
  year={2025}
}

@inproceedings{llm_golden,
  title={Large language model for verilog generation with code-structure-guided reinforcement learning},
  author={Wang, Ning and Yao, Bingkun and Zhou, Jie and Hu, Yuchen and Wang, Xi and Jiang, Zhe and Guan, Nan},
  booktitle={2025 IEEE International Conference on LLM-Aided Design (ICLAD)},
  pages={164--170},
  year={2025}
}

@INPROCEEDINGS{hlsfactory,
  author={Abi-Karam, Stefan and Sarkar, Rishov and Seigler, Allison and Lowe, Sean and Wei, Zhigang and Chen, Hanqiu and Rao, Nanditha and John, Lizy and Arora, Aman and Hao, Cong},
  booktitle={2024 ACM/IEEE 6th Symposium on Machine Learning for CAD (MLCAD)}, 
  title={{HLSFactory: A Framework Empowering High-Level Synthesis Datasets for Machine Learning and Beyond}}, 
  year={2024},
  volume={},
  number={},
  pages={1-9},
}

@inproceedings{gnn_rag,
    title = "{GNN}-{RAG}: Graph Neural Retrieval for Efficient Large Language Model Reasoning on Knowledge Graphs",
    author = "Mavromatis, Costas  and
      Karypis, George",
    booktitle = "Findings of the Association for Computational Linguistics: ACL 2025",
    year = "2025",
    publisher = "Association for Computational Linguistics",
    pages = "16682--16699",
}

@inproceedings{icrag,
    title = "Eliciting In-context Retrieval and Reasoning for Long-context Large Language Models",
    author = "Qiu, Yifu  and
      Embar, Varun R.  and
      Zhang, Yizhe  and
      Jaitly, Navdeep  and
      Cohen, Shay B  and
      Han, Benjamin",
    booktitle = "Findings of the Association for Computational Linguistics: ACL 2025",
    year = "2025",
    publisher = "Association for Computational Linguistics",
    pages = "3176--3192",
}

@article{Democratizing,
author = {Chi, Yuze and Qiao, Weikang and Sohrabizadeh, Atefeh and Wang, Jie and Cong, Jason},
title = {{Democratizing Domain-Specific Computing}},
year = {2022},
volume = {66},
number = {1},
journal = {Communications of the ACM},
pages = {74–85},
numpages = {12}
}

@article{LIFT,
  title={{LIFT: LLM-Based Pragma Insertion for HLS via GNN Supervised Fine-Tuning}},
  author={Neha Prakriya and Zijian Ding and Yizhou Sun and Jason Cong},
  journal={arXiv preprint arXiv:2504.21187},
  year={2025}
}

@article{directive_opt,
author = {Yao, Xufeng and Zhao, Wenqian and Sun, Qi and Zhuo, Cheng and Yu, Bei},
title = {High-level Synthesis Directives Design Optimization via Large Language Model},
year = {2025},
volume = {30},
number = {5},
journal = {ACM Transactions on Design Automation of Electronic Systems},
articleno = {78},
numpages = {24},
}

@article{fixme, 
title={{FIXME: Towards End-to-End Benchmarking of LLM-Aided Design Verification}}, 
volume={40}, 
number={2}, 
journal={Proceedings of the AAAI Conference on Artificial Intelligence}, 
author={Wan, Gwok-Waa and Wong, SamZaak and Su, Shengchu and Niu, Chenxu and Wang, Ning and Wan, Xinlai and Chen, Qixiang and Xing, Mengnv and Zhang, Jingyi and Ye, Jianmin and Wang, Yubo and Song, Rongchang and Ni, Tao and Xu, Qiang and Guan, Nan and Jiang, Zhe and Wang, Xi and Chen, Yong and Yang, Jun}, 
year={2026}, 
pages={1087-1095} 
}

@article{chatsva,
  title={{ChatSVA: Bridging SVA Generation for Hardware Verification via Task-Specific LLMs}},
  author={Fu, Lik Tung and Zhou, Jie and Ren, Shaokai and Zhang, Mengli and Xiong, Jia and Jiang, Hugo and Guan, Nan and Wang, Xi and Yang, Jun},
  journal={arXiv preprint arXiv:2604.02811},
  year={2026}
}

@article{iDSE,
  title={{iDSE: Navigating Design Space Exploration in High-Level Synthesis Using LLMs}},
  author={Runkai Li and Jia Xiong and Xi Wang},
  journal={arXiv preprint arXiv:2505.22086},
  year={2025}
}

\appendix

\label{sec:appendix}
\newpage

\section*{Appendix}

\section{Implementation Details of ChatHLS}

This section details how ChatHLS retrieves HLS-related context during the generation phase, parses error messages during the debugging phase, and training methods during the optimization phase.

\subsection{The Generation Stage}

The end-to-end HLS design process begins with transforming natural language or algorithms into HLS-C.
To equip the LLM with essential domain knowledge, we employ a Retrieval-Augmented Generation (RAG) technique. A specialized knowledge base is constructed from official Vitis HLS documentation (\textit{Vitis High-Level Synthesis User Guide (UG1399)}), which is segmented into chunks with a length of 1000 and an overlap of 200 between adjacent chunks to ensure contextual integrity during retrieval. When the LLM processes a critical code segment during the transformation task, it queries this embedded knowledge base. The retrieved text, detailing specific definitions, constraints, or usage patterns, is then appended to the prompt as context. This process significantly improves the success rate of the initial conversion by grounding the generation in HLS specification. 

\subsection{The Verification and Debugging Stage}
\label{appendix:debug stage}

Both the initial generation and subsequent optimization phases may introduce HLS-specific errors. These may stem from a designer's unfamiliarity with HLS constraints or from LLM-driven processes. To this end, we propose a debugging framework that relies on verification using vendor HLS tools. Current HLS tools provide detailed error messages when code issues arise, which can be leveraged to strengthen the verification process.

Initially, we collect golden results of various HLS designs from the open-source community.
These results originate from the outputs generated by HLS-C running with fixed inputs that yield correct outcomes. By applying the same inputs to the generated HLS-C and comparing its outputs against the golden results, we can verify the functional correctness of HLS-C. Additionally, beyond providing standard C compilation error messages, the HLS tool generates error feedback for HLS incompatibility errors and directive usage errors during the C-Simulation and Synthesis stages.

The debugging process identifies and addresses issues in HLS-C by first detecting errors from the parsed error message. However, the raw compilation and verification reports generated by HLS tools are extremely verbose for debugging, often spanning thousands of lines of allocation, scheduling, and binding information.
To address this, we implement a parsing mechanism that distills the essential diagnostic information. By employing regular expression matching and keyword detection, our parser identifies sentinel phrases that mark the beginning and end of each stage, as well as critical lines containing \texttt{ERROR} and \texttt{WARNING} indicators. The result is a concise, structured error message that encapsulates a clear pass/fail status for each stage, a high-level classification of the failure reason (e.g., "Synthesis time-out," "Undeclared identifier," or "Inconsistent simulation result"), and a list of the specific lines detailing the issue.

\subsection{The Optimization Stage}

To inform HLSTuner, we guide the model to perform a qualitative analysis rather than a precise quantitative prediction. By focusing on the trend of QoR changes, the LLM learns generalizable design rules. For example, consider a case where the \texttt{UNROLL} directive is inserted into a loop. The LLM compares the QoR before and after optimization. It observes that the latency decreases significantly, while the resource utilization (specifically DSP and LUTs) increases. Based on this contrast, the model generates a reasoning step explicitly linking the directive to its hardware impact: \textit{``Applying \texttt{UNROLL} increases parallelism, which reduces total execution time but requires more logic resources to instantiate parallel hardware units.''}
This approach ensures that HLSTuner learns the causal logic of optimization rather than overfitting to the specific numerical statistics of the training set. Consequently, the model can effectively generalize to unseen code structures by applying these learned hardware design principles to balance trade-offs.

\subsection{Reliability of CoT Generation}

To address potential biases in training data, we do not prompt the teacher model (DeepSeek-V3.2) to generate solutions from scratch. Instead, it is limited to explaining existing, verified results. This supervision ensures that CoT generation is based on objective data rather than internal prior knowledge.

\textbf{Optimization CoT Generation.}
For HLSTuner, we provide the teacher model with source code, initial QoR, optimized code, and optimized QoR. The LLM explains \textit{why} the provided solution leads to the performance improvements. This ensures that the reasoning is based on real performance metrics rather than LLM assumptions. Our training objective focuses on reasoning about the impact of directive combinations on QoR variations, rather than fitting absolute resource utilization and latency metrics. Hardware platform differences primarily affect the resource budget. ChatHLS avoids overfitting to a single hardware platform by learning QoR-aware reasoning and understanding the causal link between directives and performance. Our approach is constrained by utilization budgets and adapts directives based on QoR feedback.

\textbf{Debugging CoT Construction.}
Similarly, for HLSFixer, we use verified \textit{golden code} (source code) as a reference. We provide the teacher model with both the buggy code and the correct version. The model compares them to generate the diagnosis and repair steps. This prevents the model from suggesting incorrect fixes.

\section{Examples of BugRAG entries}
\label{appendix:voda}

\textit{Verification-Oriented Data Augmentation (VODA)} methodology continuously collects and categorizes common error types encountered during HLS-C generation and optimization. BugRAG contains 33 modular error slices and uses dense retrieval (all-MiniLM-L6-v2 in a Chroma vector store) with top-$k=2$, injecting only the matched slices instead of the full database. As detailed in Table \ref{hls_errors_appendix}, we have organized these errors into five principal categories that span the typical HLS design workflow: \textit{(1) HLS-C Incompatible Errors}, which involve C++ constructs not synthesizable by HLS tools, such as dynamic memory or unsupported data types.  \textit{(2) HLS-C Simulation Errors}, which arise during runtime and include issues like infinite loops or out-of-bounds array access. \textit{(3) HLS-C Compilation Errors}, covering standard C++ syntax and scope issues. \textit{(4) HLS-C Functional Errors}, where the code compiles and runs but produces incorrect results due to logical flaws. \textit{(5) HLS
Directive Errors}, related to the incorrect application or syntax of HLS pragmas. From a training perspective, curating this error repository provides structured examples that pair a specific error message and code context with a root cause analysis and a validated solution. This enables the model to learn the direct mapping between tool-specific feedback (e.g., a synthesis warning or a simulation failure) and the underlying coding mistake.

\begin{table*}[!p]
\centering
\centering
\begin{adjustbox}{width=\textwidth}
\renewcommand{\arraystretch}{2.0}
\fontsize{9pt}{10pt}\selectfont
\setlength{\tabcolsep}{1mm}
\begin{tabular}
{
>{\centering\arraybackslash}m{1.68cm}
>{\centering\arraybackslash}m{2.4cm}
>{\raggedright\arraybackslash}m{4.2cm}
>{\raggedright\arraybackslash}p{10.5cm}
}
\hline\hline 
\textbf{Category} & \textbf{Error Type} & \multicolumn{1}{c}{\textbf{Error Message}} & \multicolumn{1}{c}{\textbf{Debugging Instruction}} \\
\hline 

\multirow{4}{1.6cm}[-5mm]{\centering HLS-C\\ Incompatible Errors} &
Undefined Methods (UDM) & \textbf{Error:} \texttt{`X`} was not declared in this scope &  
\parbox[l]{10.5cm}{\textbf{Cause:} Reference to an undeclared identifier. $\bm{\Rightarrow}$ \textbf{Diagnosis:} Ensure its declaration (e.g., via \texttt{\#include} or forward declaration) precedes its use.} \\   \cline{2-4} 

& Unsupported Data Types (UDT) & \textbf{Error:} \texttt{`X`} has incomplete type and cannot be defined & 
\parbox[l]{10.5cm}{\textbf{Cause:} Utilizes non-synthesizable constructs that cannot be mapped to static hardware resources.  $\bm{\Rightarrow}$ \textbf{Diagnosis:} Replace dynamic constructs with static arrays.} \\  \cline{2-4} 

& Type Mismatch Bug (TMB) & \textbf{Error:} Ambiguous overload for \texttt{`X`} (operand types are ...) & 
\parbox[l]{10.5cm}{\textbf{Cause:} Mixed-type arithmetic between HLS arbitrary-precision and native C/C++ floating-point types.  $\bm{\Rightarrow}$ \textbf{Diagnosis:} Cast native floating-point types to the HLS type.} \\ \cline{2-4} 

& Illegal Keywords (IGK) & \textbf{Error:} \texttt{`X`} does not name a type or was not declared in scope & 
\parbox[l]{10.5cm}{\textbf{Cause:} The use of non-C/C++ keywords (e.g., \texttt{def}, \texttt{self}).  $\bm{\Rightarrow}$ Replace the invalid keywords and syntax with their standard C/C++ counterparts.} \\

\hline 

\multirow{4}{1.6cm}[-11mm]{\centering HLS-C Simulation Errors} &
Faulty Indexing (FIN) & \textbf{Error:} @E Simulation failed: nonzero return value & 
\parbox[l]{10.5cm}{\textbf{Cause:} Array access with an index outside its declared bounds. $\bm{\Rightarrow}$ \textbf{Diagnosis:}  Correct loop boundaries and ensure data-dependent indices are within valid range.} \\ \cline{2-4}

& Top Function Not Found (TFF)  & \textbf{Error:} Undefined reference to ... or a multiple definition of .... & 
\parbox[l]{10.5cm}{\textbf{Cause:} Top function name in settings mismatches the source code or the same function is defined in multiple source files. $\bm{\Rightarrow}$ \textbf{Diagnosis:} Correct the function name to match or remove the redundant source files from the project.} \\ \cline{2-4}

& Infinite Loop (INF) & \textbf{Error:} @E Simulation failed: SIGSEGV & 
\parbox[l]{10.5cm}{\textbf{Cause:} Symptomatic of a loop with a missing or logically flawed termination condition. $\bm{\Rightarrow}$ \textbf{Diagnosis:} Review the loop's exit condition to ensure it is reachable.} \\ \cline{2-4}

& Initialization Missing (INIT) & \textbf{Error:} @E Simulation failed: nonzero return value & 
\parbox[l]{10.5cm}{\textbf{Cause:} Referencing a variable that is either undeclared or uninitialized. $\bm{\Rightarrow}$ \textbf{Diagnosis:} Ensure every variable is declared within the correct scope before its first use.} \\ \cline{2-4}

& Misconfigured Loop Unit (MLU) & \textbf{Error:} @E Simulation failed: nonzero return value & 
\parbox[l]{10.5cm}{\textbf{Cause:} A logic error within the kernel code, such as incorrect loop termination conditions. $\bm{\Rightarrow}$ \textbf{Diagnosis:} Review the algorithm, particularly loop control logic.} \\ 

\hline 

\multirow{9}{1.6cm}[-10mm]{\centering HLS-C Compilation Errors} &
Illegal Comment (ICT) & \textbf{Error:} Expected ... before \texttt{`/`} token & 
\parbox[l]{10.5cm}{\textbf{Cause:} Malformed comment syntax (e.g., unclosed \texttt{/*}, incorrect \texttt{//}). $\bm{\Rightarrow}$ \textbf{Diagnosis:} Verify comment syntax and ensure no critical code has been disabled by comments.} \\ \cline{2-4} 

& Missing Colons (MCS) & \textbf{Error:} Expected \texttt{`;`} or \texttt{`:`}  before \texttt{`\}`} token &
\parbox[l]{10.5cm}{\textbf{Cause:} Missing a semicolon (\texttt{;}) for statement termination or a colon (\texttt{:}), etc. $\bm{\Rightarrow}$ \textbf{Diagnosis:} Examine the line number reported and insert the required symbol.} \\  \cline{2-4} 

& Unclosed String (UCS) & \textbf{Error:} Missing terminating \verb|'| character or \texttt{"} character & 
\parbox[l]{10.5cm}{\textbf{Cause:} An unclosed string or character literal. $\bm{\Rightarrow}$ \textbf{Diagnosis:} Locate the line reported and inspect the string and character literals.} \\ \cline{2-4} 

& Undefined Objects (UDO) & \textbf{Error:} \texttt{`X`} was not declared in this scope &
\parbox[l]{10.5cm}{\textbf{Cause:} A variable is referenced before its declaration, or its declaration is outside the accessible scope. $\bm{\Rightarrow}$ \textbf{Diagnosis:} Verify the declaration.} \\ \cline{2-4} 

& Use of Undeclared Identifier (UUI) & \textbf{Error:} \texttt{`X`} was not declared in this scope & 
\parbox[l]{10.5cm}{\textbf{Cause:} A function is referenced before its declaration. $\bm{\Rightarrow}$ \textbf{Diagnosis:} Verify the function spelling and ensure its declaration precedes any use.} \\ \cline{2-4}

& Unclosed Parentheses (UCP) & \textbf{Error:} Expected \texttt{`)`} or \texttt{`\}`} before ... token & 
\parbox[l]{10.5cm}{\textbf{Cause:} Missing parentheses (\texttt{)}) or braces (\texttt{\}}).  $\bm{\Rightarrow}$ \textbf{Diagnosis:} Ensuring code blocks (e.g., for loops, conditionals) are correctly enclosed in braces.} \\ \cline{2-4}

& Illegal Separation (ILS) & \textbf{Error:} Expected \texttt{','} or \texttt{';'} before ... token & 
\parbox[l]{10.5cm}{\textbf{Cause:} Incorrect or missing syntactic elements that disrupt code structure, such as commas.  $\bm{\Rightarrow}$ \textbf{Diagnosis:} Scrutinize the code for C++ syntax compliance.} \\ \cline{2-4}

& Head File Missing (HFM) & \textbf{Error:} \texttt{`X`} was not declared in this scope or unknown type \texttt{`X`} & 
\parbox[l]{10.5cm}{\textbf{Cause:} Missing \texttt{\#include} directive for the header file that defines the type.  $\bm{\Rightarrow}$ \textbf{Diagnosis:} Verify that all required header files, especially those defining custom or library-specific data types, are included at the beginning of the source file.} \\

\hline 

\multirow{5}{1.6cm}[-7.5mm]{\centering HLS-C Functional Errors} &
Misused Equal (MUE) & \textbf{Error:} lvalue required as left operand of assignment & 
\parbox[l]{10.5cm}{\textbf{Cause:} An assignment (\texttt{=}) was used where a comparison (\texttt{==}) was needed.  $\bm{\Rightarrow}$ \textbf{Diagnosis:} Use the correct operator for assignment (\texttt{=}) or comparison (\texttt{==}).} \\ \cline{2-4}

& Condition Error (CDE)  & \textbf{Error:} @E Simulation failed: nonzero return value. & 
\parbox[l]{10.5cm}{\textbf{Cause:} The logical checks or conditions are incorrect or misaligned with the intended logic. $\bm{\Rightarrow}$ \textbf{Diagnosis:} Trace intermediate values to find where the logic fails.} \\ \cline{2-4}

& Misaligned Zero Output (ZERO) & \textbf{Error:} @E Simulation failed: nonzero return value. & 
\parbox[l]{10.5cm}{\textbf{Cause:} A variable was accidentally initialized to zero. $\bm{\Rightarrow}$ \textbf{Diagnosis:} Focusing on variable initializations and calculations to find the mistake.} \\  \cline{2-4}

& Operation Error (OPE)  & \textbf{Error:} @E Simulation failed: nonzero return value. & 
\parbox[l]{10.5cm}{\textbf{Cause:} Mistakes in the use of operators, or misuse of functions. $\bm{\Rightarrow}$ \textbf{Diagnosis:} Trace intermediate variables based on golden results and identify the logical inconsistency.} \\ \cline{2-4}

& Bit Shift Error (SHT) & \textbf{Error:} @E Simulation failed: nonzero return value. & 
\parbox[l]{10.5cm}{\textbf{Cause:} A bit shift amount is too large for the data type, causing data loss or errors. $\bm{\Rightarrow}$ \textbf{Diagnosis:} Check the variables involved in the bit shift to find the mistake.} \\ 

\hline 

\multirow{5}{1.6cm}[-10mm]{\centering HLS Directive Errors} &
Misplaced Loop Label (MLL) & \textbf{Error:} Stray \texttt{`\#`} in program; \texttt{`pragma`} was not declared ... & 
\parbox[l]{10.5cm}{\textbf{Cause:} A C label or pragmas are incorrectly positioned relative to loop structures. $\bm{\Rightarrow}$ \textbf{Diagnosis:} Position the pragma on the line directly preceding the loop.} \\ \cline{2-4}

& Pipeline-Unroll Conflict (PUC)  & \textbf{Error:} Pragma conflict happens on \texttt{UNROLL} and \texttt{PIPELINE} & 
\parbox[l]{10.5cm}{\textbf{Cause:} Applying both \texttt{PIPELINE} and \texttt{UNROLL} to the same loop or nested loops. $\bm{\Rightarrow}$ \textbf{Diagnosis:} Inspect the loop structure indicated in the error message to find and resolve the conflicting pragmas.} \\ \cline{2-4}

& Array Partition Type (APT) &\textbf{Warning:} Unable to schedule \texttt{`load`} operation on array \texttt{`X`} due to limited memory ports & 
\parbox[l]{10.5cm}{\textbf{Cause:} Using the \texttt{PARTITION} with an inappropriate partition type (\texttt{complete}, \texttt{block}, or \texttt{cyclic}). $\bm{\Rightarrow}$ \textbf{Diagnosis:} Analyze the memory access patterns to select an appropriate partitioning \texttt{type} that provides sufficient concurrent access ports.} \\ \cline{2-4}

& Factor Not Divisible (FND) &\textbf{Warning:} Unable to schedule \texttt{`store`} operation on array \texttt{`X`} due to limited memory ports & 
\parbox[l]{10.5cm}{\textbf{Cause:} The specified partition \texttt{factor} is incongruent with the array dimensions or the loop memory access patterns. $\bm{\Rightarrow}$ \textbf{Diagnosis:} Verify that the partition \texttt{factor} is a divisor of the corresponding array dimension.} \\ \cline{2-4}

& Unknown Pragma Ignored (UNP) &\textbf{Warning:} Unknown HLS pragma ignored & 
\parbox[l]{10.5cm}{\textbf{Cause:} A pragma contains a syntax error, such as a misspelled directive. $\bm{\Rightarrow}$ \textbf{Diagnosis:} Select the correct syntax compatible with vendor HLS tools.} \\

\hline\hline
\end{tabular}
\end{adjustbox}

\caption{Examples of \textit{BugRAG} Entries}
\label{hls_errors_appendix}

\end{table*}

\begin{figure}[t]
    \centering
	\includegraphics[width=7.8cm]{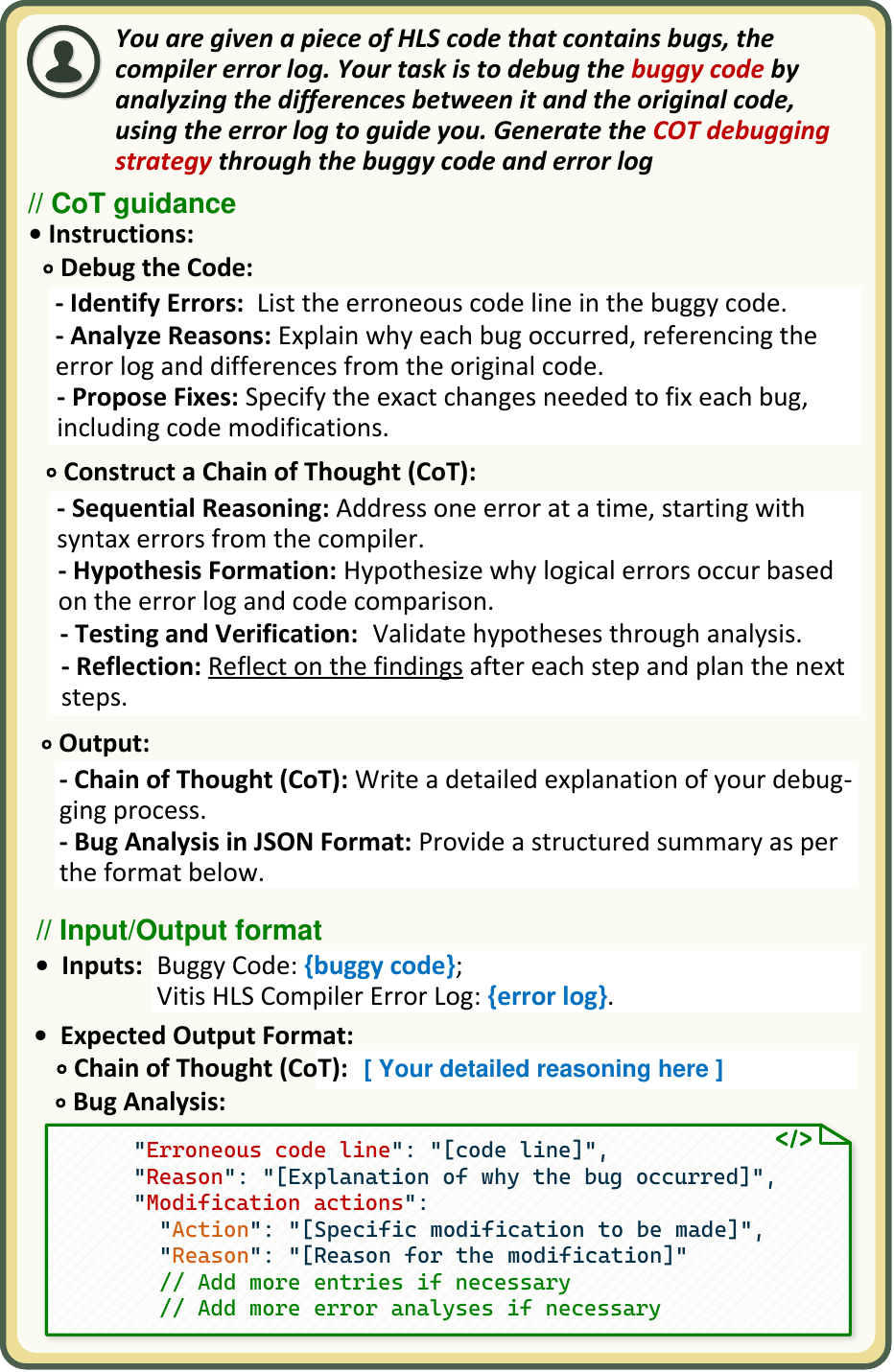} 
    \caption{Prompt for debugging analysis agent.}
	\label{prompt1}
    \vspace{-10pt}
\end{figure}

\begin{figure}[t]
    \centering
	\includegraphics[width=7.8cm]{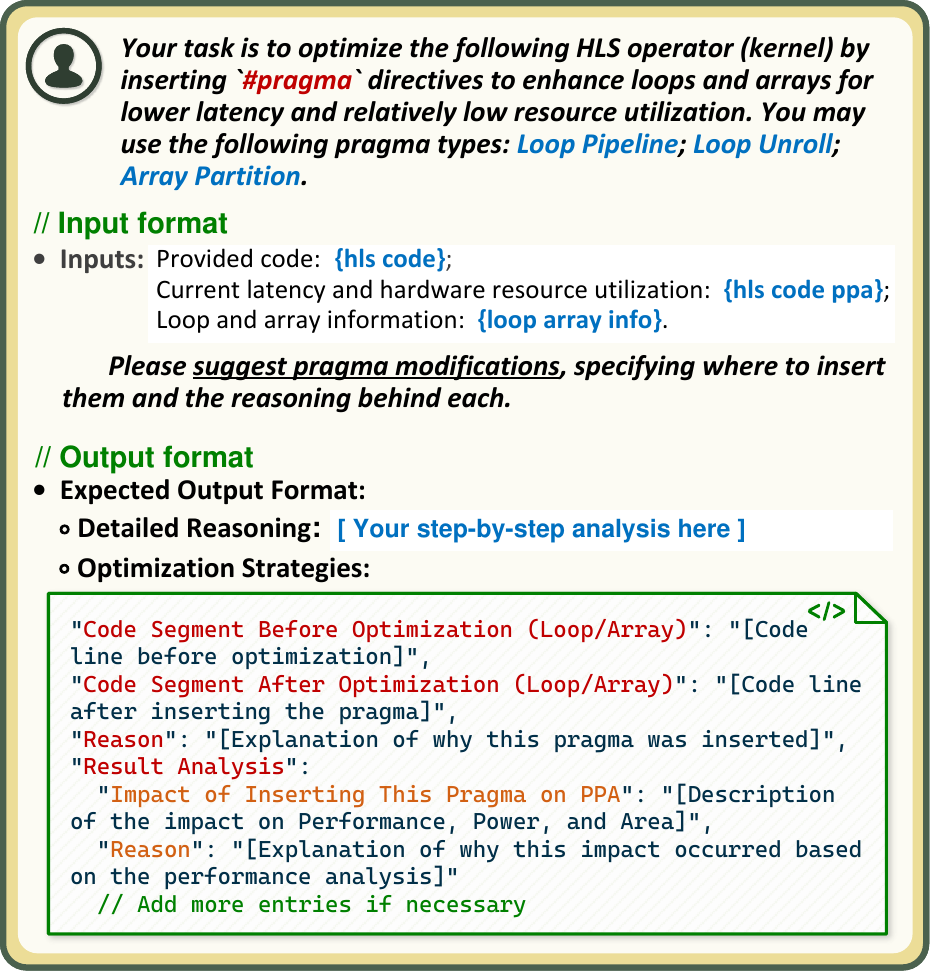} 
    \caption{Prompt for HLS design optimization agent.}
	\label{prompt2}
    \vspace{-10pt}
\end{figure}

\section{Prompt Design}

\subsection{Prompt for HLS-C Debugging Analysis}

To obtain a precise and structured debugging response from LLM, we designed a comprehensive prompt that meticulously guides its reasoning, as shown in Figure \ref{prompt1}. This \textit{reasoning-to-instruction} method enables the analysis LLM to leverage error messages to examine the error code thoroughly. The generated instructions are sequentially structured to mirror expert debugging workflows.
First, identify specific error lines by cross-referencing the code with the parsed error message. Second, formulate and validate hypotheses about the root cause of each bug, considering both syntax errors and logical inconsistencies. Finally, propose concrete code modifications.

\subsection{Prompt for HLS Design Optimization}

As illustrated in Figure \ref{prompt2}, this prompt comprises three steps: (1) define the optimization scope by specifying three target directives, (2) analyze contextual inputs including original code, performance metrics, and loop/array information to identify optimization bottlenecks, and (3) generate structured optimization strategies with explicit before-after code comparisons and performance impact assessments. This structured process closely mirrors expert-like HLS optimization, facilitating directive tuning that achieves an optimal balance between latency reduction and resource utilization constraints.

\section{Experiment Setting}
\label{appendix:setting}

\subsection{Similarity Analysis}
\label{appendix:similarity}
\begin{figure}[t]
    \centering
	\includegraphics[width=7.8cm]{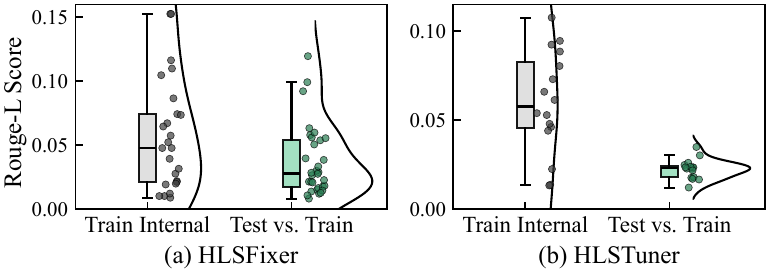} 
    \caption{Comparison of Rouge-L similarity.}
	\label{similiarity}
    \vspace{-10pt}
\end{figure}

To quantitatively evaluate the dataset similarity for our training and testing settings, HLSFixer and HLSTuner, we employed the Rouge-L metric. This metric yields a score within the $[0, 1]$ range, where smaller values correspond to lower similarity. This analysis contrasted the internal homogeneity of each SFT training set (serving as a baseline) with the similarity between individual test tasks and their corresponding training sets. The metric quantifies similarity by computing the Longest Common Subsequence (LCS) between the aggregated SFT training set ($S_{agg}$) and an individual test task ($T_i$), with the F-measure ($\beta=1$) formulated as follows:
\begin{equation}
Rouge-L = \frac{2 \cdot LCS(S_{agg}, T_{i})}{length(T_{i}) + length(S_{agg})}
\end{equation}
The results validate the efficacy of our data partitioning strategy, as depicted in Figure \ref{similiarity}. All $Rouge-L$ scores were found to be substantially below 0.15, indicating a significant dissimilarity between the training and test sets. This distinction is crucial for mitigating the risk of data leakage and ensuring a rigorous evaluation of the model's generalization. Specifically, the training set was markedly lower than its internal similarity baseline. This trend was even more pronounced in HLSTuner stage, where the mean similarity experienced a sharp decrease. This discrepancy indicates that our training set encompasses a diverse range of HLS-specific errors and provides extensive coverage of HLS-C structures, which enables our model to perform well on unseen data.

\subsection{Detailed Training Settings}
\label{appendix:training_detail}
The experiments were conducted on a server equipped with 2× Intel Xeon Platinum 8480+ CPUs and 8× NVIDIA H800-80G GPUs. The system ran Ubuntu 22.04.2 LTS with CUDA 12.2. The training, based on the Qwen-2.5-Coder-14B-Instruct model, consisted of two main stages: Supervised Fine-Tuning (SFT) followed by Direct Preference Optimization (DPO). We utilized the AdamW optimizer with bfloat16 mixed-precision.

For the SFT stage, we employed full-parameter fine-tuning, leveraging the DeepSpeed ZeRO stage 3 strategy to efficiently manage resources. The model was trained on datasets of 10,878 samples for debugging and 4,804 samples for optimization, with training durations of 72 minutes and 50 minutes, respectively. We trained for 3 epochs with a learning rate of 1e-5 under a cosine schedule with 0.1 warmup ratio, using a per-device batch size of 1 and 2 gradient accumulation steps. 

Subsequently, for the DPO stage, we initialized from the SFT model and transitioned to use Low-Rank Adaptation (LoRA) with a rank of 8 applied to all target modules. The model was trained for 2 epochs on 3,716 preference dataset took 30 min. We used a sigmoid loss function with $\beta=0.1$. The learning rate was set to 5e-6, again with a cosine scheduler and 0.1 warmup ratio. For this phase, the gradient accumulation steps were increased to 8.

\subsection{HLSFixer Setting}
\label{appendix:HLSFixer_Setting}

\textbf{HLS-C Generation Task.}
We used 108 design tasks, covering 85 designs from HLS-Eval \cite{hlseval} and 23 additional custom designs. This benchmark includes scientific computing (PolyBench), embedded systems (MachSuite), and cryptography (CHStone) to ensure broad representativeness. In Table \ref{table:generation}, we compare DeepSeek-V3.2 and Gemini-3-pro. Both models are accessed via API with the temperature set to 0.7. In Figure \ref{c2hlsc}, baseline results for HLSRewriter \cite{hlsrewriter} and C2HLSC \cite{C2HLSC} are derived from their reported transformation pass rates (defined as pass@1 over 10 attempts).

\textbf{HLS-C Debugging Task.}
We constructed 591 debugging test cases derived from 32 correct HLS designs. We used automated error injection techniques to introduce the 34 errors compiled in Table \ref{hls_errors_appendix}. We parsed Xilinx Vitis HLS compilation reports and only retained the cases where the tool explicitly reported the type of injected error. This ensures that the specific buggy samples are unique and unseen during the model pretraining or fine-tuning.
In the debugging experiments, the temperature for the analysis model is set to 0.7. For the code modification model, we use DeepSeek-V3.2 and set its temperature to 0.1 to ensure strict instruction adherence. To evaluate the debugging capability of HLSFixer, we keep the code modification model fixed and change the analysis model for comparison.
Each benchmark is equipped with a golden testbench to verify the functional correctness. A debugging attempt is considered successful if the corrected code passes CSIM, CSYN, and COSIM.

\subsection{HLSTuner Baseline Settings and Metrics}
\label{appendix:HLSTuner}

\begin{table}[t]
    \centering
    \fontsize{9pt}{10pt}\selectfont
    \renewcommand{\arraystretch}{1.2}
    \begin{adjustbox}{width=0.47\textwidth}
    \begin{tabular}{cccccc}
        \hline\hline %
         \textbf{Kernel} & \textbf{Atax}
         & \textbf{Bicg} & \textbf{Gemm} & \textbf{Gesummv} & \textbf{Mvt} \\
        \hline
        Lat. (Cycles) & 1702 & 1658 & 15661 & 470 & 1629 \\ 
        DSP (Util.) & 14.6\% & 13.2\% & 10.5\% & 10.8\% & 13.9\% \\ 
        FF (Util.) & 1.6\% & 1.6\% & 0.4\% & 0.8\% & 1.7\% \\ 
        LUT (Util.) & 3.6\% & 3.3\% & 2.1\% & 2.4\% & 3.8\% \\
        \hline
        Loop \& Array & 4 / 4 & 3 / 5 & 4 / 3 & 2 / 5 & 4 / 5 \\ 
        \# Directives & 20 & 21 & 17 & 19 & 23 \\
        \hline\hline 
    \end{tabular}
    \end{adjustbox}

    \caption{QoR metrics and design structure of representative baseline computation kernels in linear algebra.}
    \label{Synthesis Results}    
    \vspace{-10pt}
\end{table}

\textbf{Vitis HLS Auto-Optimization Baseline.} 
The HLS tool applied default pipeline optimizations to the loops in the given design, which causes the baseline test results to generally be better than those without any optimization. The synthesis results of the HLS kernels selected in the optimization task are shown in Table \ref{Synthesis Results}, including loop/array information, QoR (latency and resource utilization).

\textbf{Optimization Objective.}
Given the strict performance and throughput requirements of compute-intensive operations, our evaluation specifically focuses on minimizing latency while maintaining hardware-constrained resource efficiency to validate the effectiveness. By imposing fixed hardware constraints on resource utilization, we explored optimization schemes within a reasonable design space to ensure effective design. To validate the efficiency of HLSTuner, we evaluated the speedup ratios achieved through fewer than five optimization iterations, demonstrating its ability to generate effective solutions under limited search budgets.

\begin{table}[t]
\normalsize
\centering
\fontsize{10pt}{10pt}\selectfont
\renewcommand{\arraystretch}{1.3}
\begin{adjustbox}{width=0.47\textwidth}
\begin{tabular}{cccc}
\hline\hline
Structure & Optimization Directive & Configuration \\ \midrule
\multirow{2}{*}{Loop} & \texttt{PIPELINE} &  ``off",  ``on" \\
 & \texttt{UNROLL} & integer & \\
\hline
\multirow{2}{*}{Array}& \multirow{2}{*} {\texttt{ARRAY\_PARTITION}} & ``complete",  ``block",  ``cyclic" \\
 & & integer& \\ 
 \hline\hline
\end{tabular}
\end{adjustbox}
\caption{HLS directive type and configuration.}
\label{pragma_config}
\vspace{-10pt}
\end{table}

\textbf{Performance Metrics for HLSTuner.}
In the context of HLS design evaluation, the latency is derived from synthesis timing analysis rather than post-implementation routing delays. This metric captures circuit performance determined by logic-level optimization decisions, eliminating variations introduced by layout and wiring during implementation. 
For an HLS design $\lambda(\theta)$, $\theta$ represents the inserted HLS directives. Under vendor HLS tool $\mathcal{H}$, we analyze the QoR of the explored design using latency $Lat(\mathcal{H}, \lambda(\theta))$ and resource utilization $Util(\mathcal{H}, \lambda(\theta))$.
To quantify the impact of our optimization strategies, we measure the $Speedup$:
\begin{equation}
\label{eq:speedup_pragma}
\underbrace{Speedup = \frac{Lat(\mathcal{H}, \lambda(\theta_{baseline}))}{Lat(\mathcal{H}, \lambda(\theta_{optimized}))}}_{Util_r(\lambda) \leq 80\%, \forall r \in \{DSP, FF, LUT\}}
\end{equation}
This metric reflects the performance improvement attributable to the chosen optimization directives $ \theta_{\text{optimized}} $ for a given design $ \lambda $.

\textbf{Experimental Setup.}
We set the synthesis time limit to 1 hour. Excessively high parallelism configurations cause the scheduling and binding phases of vendor HLS tool to spend more time adjusting the Initiation Interval (II). Exceeding this time limit will likely result in resource over-utilization. We consider this to be a failed optimization attempt. If all 15 attempts fail, or if the optimized latency falls below the baseline, or if resource utilization exceeds limits, we set its penalty speedup in Figure \ref{fig:performance}. For example, DeepSeek-V3.2 yields no valid optimization across all 15 attempts in \textit{gemm}, \textit{2mm}, \textit{symm}, \textit{syrk} and \textit{gemm\_ncubed}, we set its speedup to 1 to penalize its geo mean result.

\subsection{HLS Directives Supported by HLSTuner.}

The optimization of loop and array parallelism constitutes a critical bottleneck in HLS design, as these structural elements predominantly determine the performance characteristics of the synthesized hardware implementation.
Specifically, for loops, HLSTuner supports the pragma \texttt{PIPELINE} and \texttt{UNROLL}. The \texttt{PIPELINE} pragma allows overlapping execution of loop iterations to improve throughput, while the \texttt{UNROLL} pragma replicates loop bodies to exploit parallelism. For arrays, HLSTuner supports \texttt{ARRAY\_PARTITION} pragma, which divides arrays into smaller memories to enable parallel access and reduce memory bottlenecks. Table \ref{pragma_config} summarizes these optimization directive configurations.

\section{RAG Baselines}
\label{app:rag_fairness}

To address the concern that the comparisons against general-purpose LLMs may be unfair without retrieval augmentation, we conduct a controlled study on \textit{matched RAG baselines} for both HLS-C generation and debugging. In all experiments, we use identical task descriptions, identical prompts, and the same retrieved context for all models whenever RAG is enabled. This setting is to isolate the contribution of retrieval from that of our specialized data curation, model adaptation, and hierarchical feedback-driven reasoning.

\begin{table}[t]
\centering
\small
\setlength{\tabcolsep}{7pt}
\begin{adjustbox}{width=0.47\textwidth}
\begin{tabular}{lccc}
\toprule
Model & CSIM & CSYN & COSIM \\
\midrule
DeepSeek-V3.2 & 47.0\% & 43.2\% & 31.5\% \\
DeepSeek-V3.2 + RAG & 42.6\% & 35.8\% & 32.1\% \\
Gemini-3-pro & 57.9\% & 56.5\% & 48.1\% \\
Gemini-3-pro + RAG & 57.7\% & 57.4\% & 50.9\% \\
ChatHLS (w/HLSFixer) & \textbf{82.1\%} & \textbf{81.2\%} & \textbf{77.2\%} \\
\bottomrule
\end{tabular}
\end{adjustbox}
\caption{Generation pass@1 under matched RAG baselines. RAG uses retrieved context from the official Vitis HLS documentation.}
\label{tab:rag_generation}
\vspace{-10pt}
\end{table}

\begin{figure*}[!t]
	\centering 
    \includegraphics[width=\textwidth]{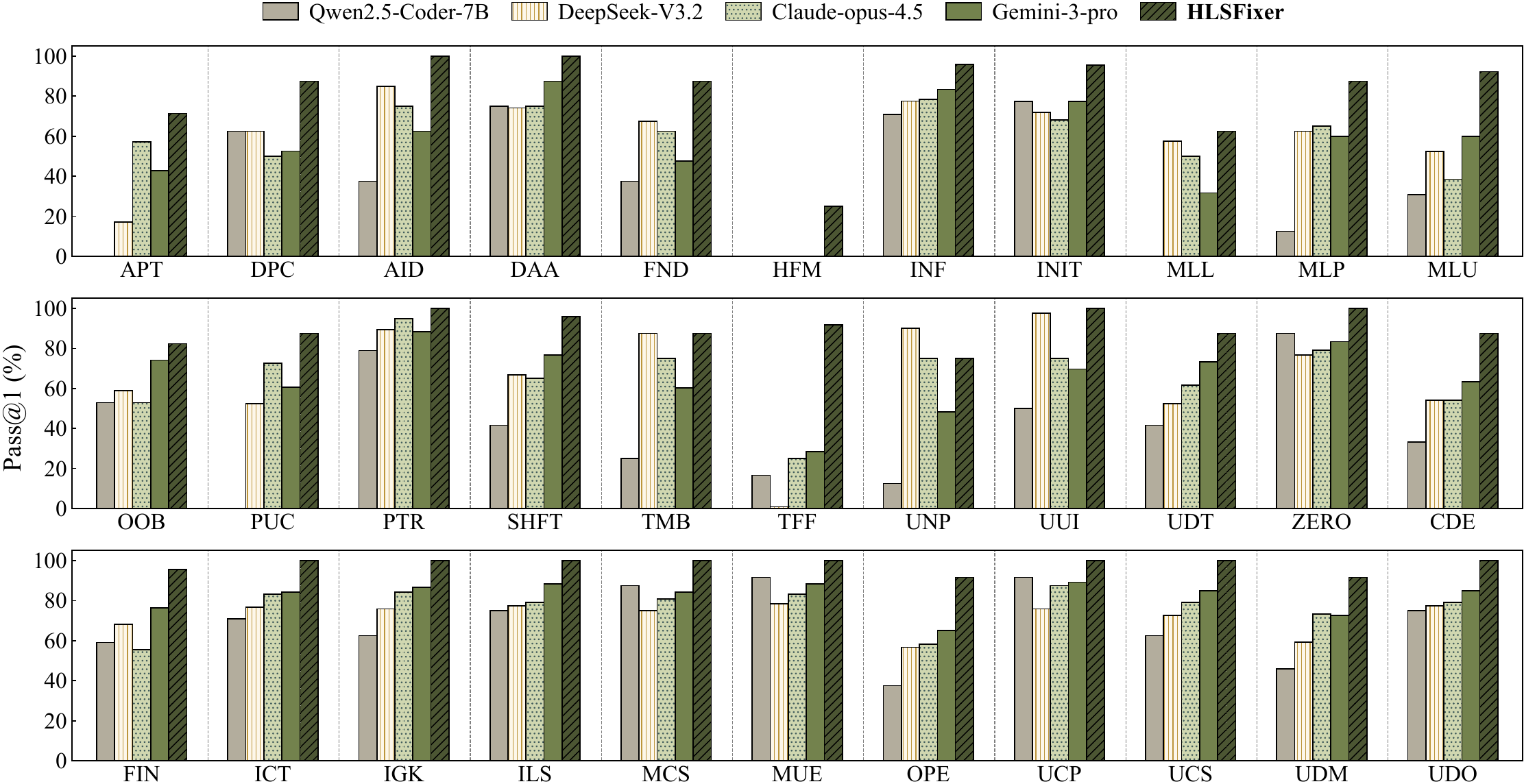} 
	\caption{Comparison of code repair pass rates on different HLS-specific errors.}
	\label{Performance}
    \vspace{-10pt}
\end{figure*}

\subsection{RAG Baselines for HLS-C Generation}
\label{app:rag_generation}

For HLS-C generation, we augment general LLMs with retrieved context from the official Vitis HLS documentation. This setting is intended to test whether direct access to domain documentation alone is sufficient to close the performance gap to ChatHLS. Table~\ref{tab:rag_generation} shows that adding documentation retrieval alone yields only marginal gains, and in some cases even degrades early-stage correctness. For DeepSeek-V3.2, RAG decreases CSIM from 47.0\% to 42.6\% and CSYN from 43.2\% to 35.8\%, while only slightly improving COSIM from 31.5\% to 32.1\%. For Gemini-3-pro, the effect is also limited. In contrast, ChatHLS substantially outperforms the strongest RAG baseline. These results suggest that access to domain documents is helpful but insufficient. In HLS-C generation, the main bottleneck is not simply missing reference material, but the lack of HLS-specific reasoning patterns for synthesizability, directive semantics, and feedback-aware correction.

\begin{table}[t]
\centering
\small
\setlength{\tabcolsep}{6pt}
\begin{adjustbox}{width=0.47\textwidth}
\begin{tabular}{lcccc}
\toprule
Model & Kernel & Vitis & Manual & Overall \\
\midrule
DeepSeek-V3.2 & 62.7\% & 40.1\% & 84.0\% & 66.3\% \\
DeepSeek-V3.2 + RAG-1 & 67.8\% & 74.2\% & 85.4\% & 79.2\% \\
DeepSeek-V3.2 + RAG-2 & 68.9\% & 74.7\% & 84.1\% & 78.9\% \\
Gemini-3-pro & 47.7\% & 69.7\% & 83.0\% & 70.4\% \\
Gemini-3-pro + RAG-1 & 55.6\% & 80.4\% & 93.9\% & 84.5\% \\
Gemini-3-pro + RAG-2 & 60.0\% & 81.4\% & 91.6\% & 83.6\% \\
HLSFixer & \textbf{78.9\%} & \textbf{95.4\%} & \textbf{96.4\%} & \textbf{93.4\%} \\
\bottomrule
\end{tabular}
\end{adjustbox}
\caption{Debugging \textit{pass@1} under RAG baselines. RAG-1 uses Vitis HLS documentation and BugRAG, RAG-2 uses the full training set as the retrieval corpus.}
\label{tab:rag_debugging}
\vspace{-10pt}
\end{table}

\subsection{RAG Baselines for HLS Debugging}
\label{app:rag_debugging}

We further evaluate matched RAG baselines for debugging under two retrieval settings, as shown in Table \ref{tab:rag_debugging}. RAG-1 represents a practical retrieval setup with compact domain knowledge and structured error slices, whereas RAG-2 tests whether scaling the retrieval corpus with the full training set can replace model adaptation.

Unlike generation, debugging benefits more substantially from retrieval augmentation. DeepSeek-V3.2 improves from 66.3\% to 79.2\% overall under RAG-1, while Gemini-3-pro improves from 70.4\% to 84.5\%. This indicates that explicit retrieval is indeed useful for grounding error analysis, especially when the model can directly map retrieved examples to error logs and local code regions. However, HLSFixer still achieves the best overall pass rate at 93.4\%, outperforming the strongest RAG baseline. This result shows that retrieval helps general-purpose LLMs, but it does not replace the need for specialized training. Our fine-tuned analysis model internalizes the hierarchical logic of error diagnosis, which leads to more stable performance across heterogeneous error sources.

Another observation is that more retrieved data does not necessarily yield better results. For both DeepSeek-V3.2 and Gemini-3-pro, RAG-2 performs slightly worse than RAG-1 in overall pass rate. This suggests that simply retrieving from the full training set can introduce weakly matched or noisy contexts, which may dilute the relevance of the evidence and interfere with the model reasoning. Therefore, the main advantage of ChatHLS is not retrieval volume, but the combination of structured error knowledge, targeted analysis supervision, and feedback-grounded reasoning.

\vspace{-5pt}

\section{HLSFixer Supplementary Results}
\label{appendix:hlsfixer_result}

\subsection{HLSFixer Error Diagnosis Reasoning}

HLSFixer demonstrates systematic error diagnosis and correction through its hierarchical analysis workflow, exemplified by two representative HLS design errors: dynamic memory allocation violations and dataflow pragma conflicts.

\textbf{Dynamic Array Allocation.}
A critical HLS design transformation challenge arises from dynamic memory allocation patterns. When synthesizing code containing C++ dynamic allocation, HLSFixer identifies unrecognized memory operators breaking synthesis flow, and unbound memory references contradicting hardware-resource preallocation principles. HLSFixer replaces dynamic allocation with static declaration to ensure deterministic memory footprint while preserving the original algorithmic intent. 

\textbf{Dataflow Pragma Conflict.}
Additionally, HLSFixer demonstrates its capability to validate and correct optimization directives by resolving non-canonical dataflow region conflicts. An incorrectly applied \texttt{DATAFLOW} to logically interdependent loops resulted in scheduling failures due to unmanaged producer-consumer dependencies. By recognizing the semantic contradiction between the dataflow pragma requirements for independent processes and the sequential dependencies of actual loop structures, HLSFixer resolves the conflicts through targeted pragma removal or pipeline stage reconstruction, ensuring valid task-level parallelism aligned with HLS scheduling semantics.

\begin{figure}[t]
    \centering
	\includegraphics[width=7.8cm]{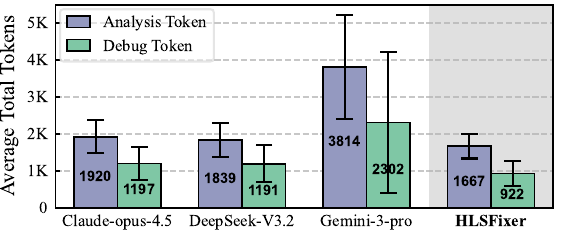} 
    \caption{Token consumption for debugging agents.}
	\label{debug_token}
\end{figure}

\begin{figure}[t]
    \centering
	\includegraphics[width=7.8cm]{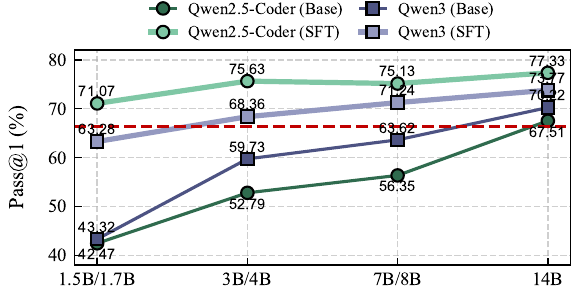} 
    \caption{Performance scaling of fine-tuned LLMs on HLS-C debugging tasks. (The red dashed line represents the Deepseek-V3.2 Pass@1 of 66.3\%.)}
	\label{scale}
    \vspace{-10pt}
\end{figure}

\subsection{Error Correction Results}

Figure \ref{Performance} compares the pass rates of HLSFixer against baseline models across 33 error types. The results demonstrate that HLSFixer significantly outperforms general LLMs, particularly in addressing HLS-specific programming constraints. For compatibility errors, HLSFixer successfully resolves Dynamic Array Allocation (DAA) and Pointer Access Error (PTR). This indicates that HLSFixer learned to avoid standard C/C++ constructs and acquired hardware-oriented coding capabilities during training. Furthermore, HLSFixer exhibits superior performance in resolving HLS directive errors, effectively fixing complex optimization issues like APT and PUC.

In Figure \ref{debug_token}, we report the token consumption of the error analysis and code modification models used in the debug task. Gemini-3-pro consumes more tokens because it involves a reasoning process. In most cases, HLSFixer can resolve errors that arise during HLS generation and optimization tasks in a single attempt. Since scenarios requiring multifaceted evaluation to resolve complex problems are infrequent, the overall token cost in the end-to-end workflow is consequently lower.

\begin{figure}[t]
    \centering
	\includegraphics[width=7.8cm]{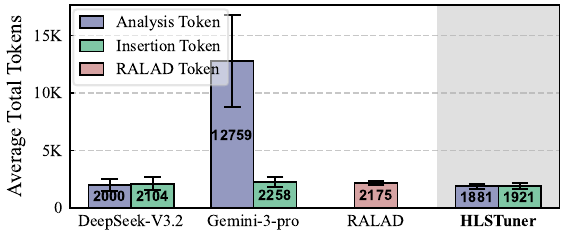} 
    \caption{Token consumption for optimization agents.}
	\label{opt_token}
    \vspace{-10pt}
\end{figure}

\begin{figure*}[!t]
	\centering 
\includegraphics[width=\textwidth]{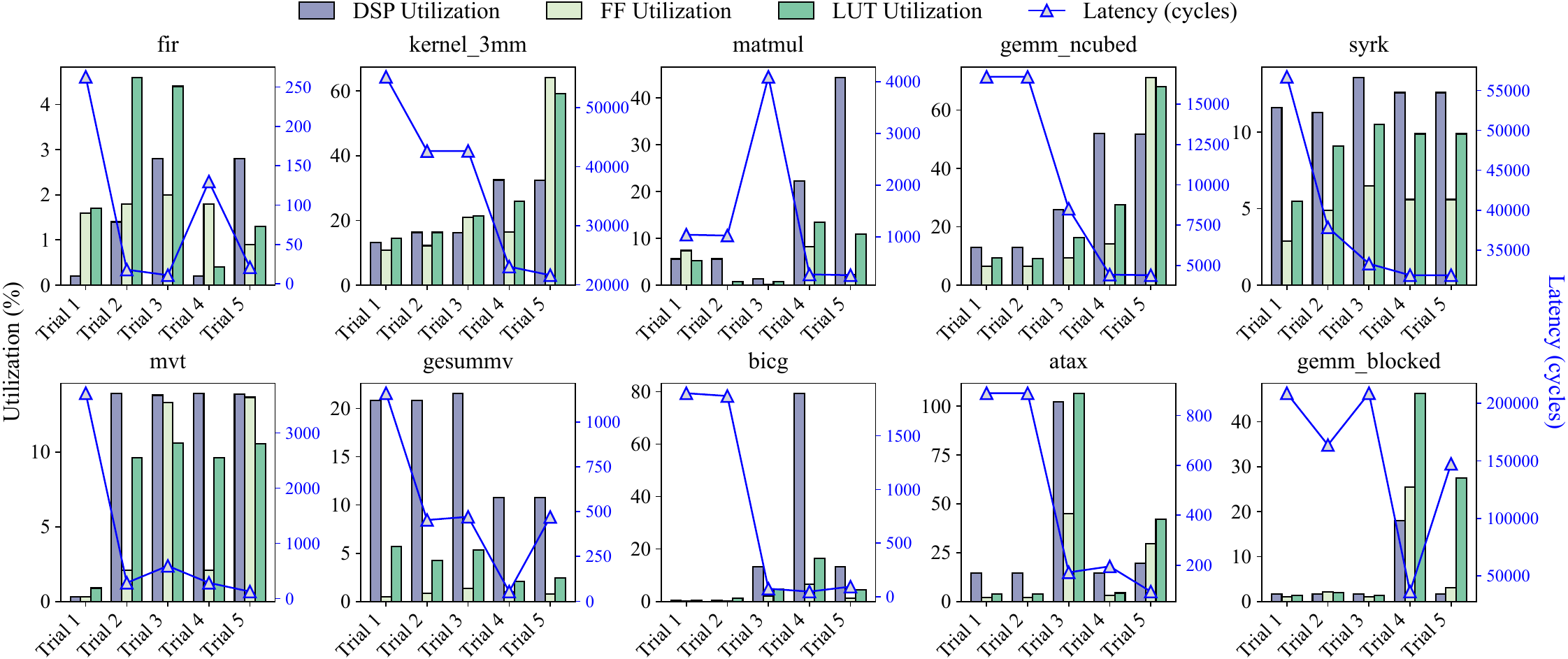} 
	\caption{Optimization trajectories HLSTuner within five optimization iterations. By analyzing QoR and dynamically tuning directives, LLM progressively maximizes performance while adhering to resource constraints.}
	\label{fig:trajectory}
    \vspace{-5pt}
\end{figure*}

\subsection{Impact of Model Scale}
\label{appendix:Scale}

To evaluate the impact of model scale and domain-specific fine-tuning on the HLS-C debugging task, we benchmarked the debugging performance of the code-centric Qwen2.5-Coder against the general-purpose Qwen3 series. To ensure a fair comparison, we employed identical training data and settings during SFT for all base models. 

As illustrated in Figure \ref{scale}, for both Qwen2.5-Coder and Qwen3, the debugging performance scales consistently with LLM parameters. Fine-tuning with the expertise required for HLS coding constraints has comprehensively enhanced debugging capabilities across all models. Notably, this process enables the base model to approach or surpass the performance of DeepSeek-V3.2. Furthermore, we observe that Qwen2.5-Coder achieves greater performance gains through SFT compared to Qwen3. We attribute this advantage to its extensive pre-training on massive code corpora. This specialized foundation grounds the fine-tuning, enabling the LLM to absorb the subtle nuances specific to HLS debugging tasks more efficiently.

\section{HLSTuner Supplementary Results}
\label{appendix:hlstuner}

\subsection{Analysis of Optimization Trajectories}

Our analysis reveals a significant performance disparity in DeepSeek-V3.2, which performs well on PolyBench but fails to generalize to other benchmarks. This issue is particularly evident with the compute-intensive kernels \textit{kernel\_2mm} and \textit{kernel\_symm}. For these kernels, DeepSeek-V3.2 consistently applied overly aggressive parallelization strategy, resulting in synthesis failures or excessive resource consumption across all 15 optimization attempts. This suggests that it has trivially memorized specific optimal configurations, likely due to data contamination, rather than genuinely learning a holistic tuning policy. 

In contrast, Gemini-3-pro and HLSTuner dynamically adjust directive tuning strategies based on the QoR feedback from each iteration. As illustrated by the optimization trajectories in Figure \ref{fig:trajectory}, HLSTuner progressively increases parallelism for performance gains or reduces it when resource utilization exceeds the specified limit. This dynamic navigation of the design space enables them to achieve sustained performance improvements while adhering to resource constraints.

In the experiment, we set DeepSeek-V3.2 as the insertion agent. Following the generated optimization strategy, including detailed directive combinations, configurations, and placements, this agent inserts HLS instructions into the source code. The average token consumption is shown in Figure \ref{opt_token}, where Gemini-3-pro consumed a large number of tokens during reasoning due to the complexity of the optimization task. For RALAD, we combined the retrieved guidance to insert directives into the source code to reproduce the original RAG settings. In contrast, our optimization strategy focused the LLM on analyzing the QoR and specific HLS directives, achieving superior optimization results.

\begin{table}[t]
    \normalsize
    \centering
    \fontsize{10pt}{10pt}\selectfont
    \renewcommand{\arraystretch}{1.2}
    \begin{adjustbox}{width=0.47\textwidth}
    \begin{tabular}{c|c|ccccc}
        \toprule
        \multicolumn{2}{c|}{\textbf{Kernel}} & \textbf{Atax} & \textbf{Bicg} & \textbf{Gemm} & \textbf{Gesummv} & \textbf{Mvt}~ \\
        \midrule
        Dahlia & \makecell[c]{
        DSP \\
        FF\\ 
        LUT} 
        & \makecell[c]{14.6\%\\1.8\%\\3.0\%}& \makecell[c]{13.5\%\\1.8\%\\2.9\%} & \makecell[c]{10.5\%\\0.4\%\\1.7\%} & \makecell[c]{10.8\%\\0.9\%\\1.9\%} & \makecell[c]{13.9\%\\2.1\%\\5.1\%}~ \\
        \midrule
        HeteroCL & \makecell[c]{
        DSP \\
        FF\\ 
        LUT} 
        & \makecell[c]{1.4\%\\0.1\%\\0.5\%}& \makecell[c]{20.1\%\\1.3\%\\10.6\%} & \makecell[c]{4.5\%\\0.4\%\\1.9\%} & \makecell[c]{10.6\%\\0.4\%\\2.1\%} & \makecell[c]{13.9\%\\0.8\%\\2.6\%}~ \\
        \midrule
        Allo & \makecell[c]{
        DSP \\
        FF\\ 
        LUT} 
        & \makecell[c]{1.4\%\\0.1\%\\0.7\%}& \makecell[c]{13.9\%\\1.1\%\\2.8\%} & \makecell[c]{4.5\%\\1.4\%\\3.1\%} & \makecell[c]{7.2\%\\3.4\%\\5.1\%} & \makecell[c]{13.9\%\\0.1\%\\1.9\%}~ \\
        \midrule
        \makecell[c]{HLSTuner\\ (Our work)}& \makecell[c]{
        DSP \\ 
        FF\\ 
        LUT} 
        & \makecell[c]{19.4\%\\29.5\%\\42.0\%}& \makecell[c]{79.2\%\\6.6\%\\16.5\%} & \makecell[c]{49.9\%\\1.2\%\\6.9\%} & \makecell[c]{10.8\%\\1.0\%\\2.1\%} & \makecell[c]{13.9\%\\13.7\%\\10.5\%}~ \\ 
        \bottomrule
    \end{tabular}
    \end{adjustbox}
    \caption{Hardware cost of optimized HLS design.}
    \label{Optimization Cost Ratio}
    \vspace{-10pt}
\end{table}

\subsection{Analysis of DSL-based and DSE Methods}

To ensure a fair comparison with DSL methods (Dahlia, HeteroCL, Allo) and HGBO-DSE, we selected PolyBench kernels with identical design scales. All evaluations were conducted using consistent HLS tool versions, synthesis time limits, and target FPGA platforms. Table \ref{Optimization Cost Ratio} summarizes the resource utilization. For \textit{Mvt} and \textit{Gesummv}, our method achieves DSP usage parity with Dahlia. For \textit{Gemm} and \textit{Bicg}, HLSTuner leverages more resources, demonstrating its ability to strategically implement aggressive parallelization strategies that fully utilize the target hardware budget.

Dahlia focuses primarily on memory banking and loop unrolling factor alignment, but lacks loop pipelining capabilities and has limited optimization options for compute-intensive applications. Allo and HeteroCL both require manual specification of unroll factors for each compute node and rely on expert knowledge to navigate the design space effectively. This reliance on hardware expertise creates significant usability barriers to achieving optimal performance. In contrast, HLSTuner eliminates the need for source code modification, achieving performance goals through automated optimization directive tuning. Although this approach does not achieve the same high efficiency in target hardware resource utilization compared to methods like Allo, it reduces developer effort by allowing brief descriptive sentences to prompt optimization.

The HGBO-DSE method based on Bayesian optimization still struggles to determine and tune critical HLS directives even after 100 iterations. This indicates that without human guidance, such methods tend to converge rapidly to suboptimal solutions in complex HLS designs. In contrast, HLSTuner identifies superior solutions within a short iteration period, as it inherently learns from expert optimization experience. This simplifies the time-consuming design space exploration process, allowing developers to focus on adjusting algorithmic structures to adapt to rapidly evolving application requirements.

\section{Information of Assets}

We strictly adhere to the licenses (MIT, Apache 2.0) and terms of use for all existing artifacts utilized in this work. Our use of these open-source assets is consistent with their intended purposes. For the artifacts we created, including fine-tuned LLMs, we specify that they are intended exclusively for scientific research. This usage is compatible with the original access conditions of the source data. 

The Repo. of code is available at: \url{https://github.com/GEAR-SEU/ChatHLS-ACL-26}

\end{document}